\documentclass[aps,prd,twocolumn,superscriptaddress,nofootinbib]{revtex4-2}
\usepackage{graphicx}  
\usepackage{dcolumn}   
\usepackage{bm}        
\usepackage{amssymb}   
\usepackage[caption=false]{subfig}
\usepackage{xcolor}
\usepackage{enumitem}
\usepackage{amsmath}

\newcommand{\minerva}{MINERvA}

\newcommand{\tune}{\minerva~Tune~v1}

\newcommand{\pt}{$p_{t}$}
\newcommand{\pz}{$p_{||}$}

\begin{document}

\title{Measurement of inclusive charged-current $\nu_\mu$ cross sections as a function of muon kinematics at $<E_\nu>\sim6~GeV$ on hydrocarbon }

\newcommand{\Rutgers}{Rutgers, The State University of New Jersey, Piscataway, New Jersey 08854, USA}
\newcommand{\Hampton}{Hampton University, Dept. of Physics, Hampton, VA 23668, USA}
\newcommand{\Dortmund}{Institute of Physics, Dortmund University, 44221, Germany }
\newcommand{\Otterbein}{Department of Physics, Otterbein University, 1 South Grove Street, Westerville, OH, 43081 USA}
\newcommand{\JMU}{James Madison University, Harrisonburg, Virginia 22807, USA}
\newcommand{\Florida}{University of Florida, Department of Physics, Gainesville, FL 32611}
\newcommand{\UCIrvine}{Department of Physics and Astronomy, University of California, Irvine, Irvine, California 92697-4575, USA}
\newcommand{\CBPF}{Centro Brasileiro de Pesquisas F\'{i}sicas, Rua Dr. Xavier Sigaud 150, Urca, Rio de Janeiro, Rio de Janeiro, 22290-180, Brazil}
\newcommand{\PUCP}{Secci\'{o}n F\'{i}sica, Departamento de Ciencias, Pontificia Universidad Cat\'{o}lica del Per\'{u}, Apartado 1761, Lima, Per\'{u}}
\newcommand{\INRM}{Institute for Nuclear Research of the Russian Academy of Sciences, 117312 Moscow, Russia}
\newcommand{\Jlab}{Jefferson Lab, 12000 Jefferson Avenue, Newport News, VA 23606, USA}
\newcommand{\Pittsburgh}{Department of Physics and Astronomy, University of Pittsburgh, Pittsburgh, Pennsylvania 15260, USA}
\newcommand{\Guanajuato}{Campus Le\'{o}n y Campus Guanajuato, Universidad de Guanajuato, Lascurain de Retana No. 5, Colonia Centro, Guanajuato 36000, Guanajuato M\'{e}xico.}
\newcommand{\Athens}{Department of Physics, University of Athens, GR-15771 Athens, Greece}
\newcommand{\Tufts}{Physics Department, Tufts University, Medford, Massachusetts 02155, USA}
\newcommand{\WM}{Department of Physics, William \& Mary, Williamsburg, Virginia 23187, USA}
\newcommand{\FNAL}{Fermi National Accelerator Laboratory, Batavia, Illinois 60510, USA}
\newcommand{\Purdue}{Department of Chemistry and Physics, Purdue University Calumet, Hammond, Indiana 46323, USA}
\newcommand{\MCLA}{Massachusetts College of Liberal Arts, 375 Church Street, North Adams, MA 01247}
\newcommand{\UMD}{Department of Physics, University of Minnesota -- Duluth, Duluth, Minnesota 55812, USA}
\newcommand{\Northwestern}{Northwestern University, Evanston, Illinois 60208}
\newcommand{\UNI}{Facultad de Ciencias, Universidad Nacional de Ingenier\'{i}a, Apartado 31139, Lima, Per\'{u}}
\newcommand{\Rochester}{University of Rochester, Rochester, New York 14627 USA}
\newcommand{\Austin}{Department of Physics, University of Texas, 1 University Station, Austin, Texas 78712, USA}
\newcommand{\USM}{Departamento de F\'{i}sica, Universidad T\'{e}cnica Federico Santa Mar\'{i}a, Avenida Espa\~{n}a 1680 Casilla 110-V, Valpara\'{i}so, Chile}
\newcommand{\Geneva}{University of Geneva, 1211 Geneva 4, Switzerland}
\newcommand{\Chicago}{Enrico Fermi Institute, University of Chicago, Chicago, IL 60637 USA}
\newcommand{\hired}{}
\newcommand{\OregonState}{Department of Physics, Oregon State University, Corvallis, Oregon 97331, USA}
\newcommand{\oxford}{Oxford University, Department of Physics, Oxford, OX1 3PJ United Kingdom}
\newcommand{\umiss}{University of Mississippi, Oxford, Mississippi 38677, USA}
\newcommand{\upenn}{Department of Physics and Astronomy, University of Pennsylvania, Philadelphia, PA 19104}
\newcommand{\AMU}{AMU Campus, Aligarh, Uttar Pradesh 202001, India}
\newcommand{\wroclaw}{University of Wroclaw, plac Uniwersytecki 1, 50-137 Wroa\l{}aw, Poland}
\newcommand{\Mohali}{Department of Physical Sciences, IISER Mohali, Knowledge City, SAS Nagar, Mohali - 140306, Punjab, India}
\newcommand{\CINVESTAV}{Departamento de Fisica Col. San Pedro Zacatenco, 07360 Mexico, DF, Av. Instituto PolitÃ©cnico Nacional, Mexico}
\newcommand{\york}{York University, Department of Physics and Astronomy, Toronto, Ontario, M3J 1P3 Canada}
\newcommand{\ND}{Department of Physics, University of Notre Dame, Notre Dame, Indiana 46556, USA}
\newcommand{\ICL}{The Blackett Laboratory,  Imperial College London,  London SW7 2BW, United Kingdom}

\newcommand{\mateusfcarneiroThanks}{Now at Brookhaven National Laboratory}
\newcommand{\emilymaherThanks}{Department of Physics}
\newcommand{\bamThanks}{Now at University of Minnesota}

\author{D.~Ruterbories}                   \affiliation{\Rochester}
\author{A.~Filkins}                       \affiliation{\WM}

\author{Z.~~Ahmad~Dar}                    \affiliation{\WM}  \affiliation{\AMU}
\author{F.~Akbar}                         \affiliation{\AMU}
\author{D.A.~Andrade}                     \affiliation{\Guanajuato}
\author{M.~V.~Ascencio}                   \affiliation{\PUCP}
\author{A.~Bashyal}                       \affiliation{\OregonState}
\author{L.~Bellantoni}                    \affiliation{\FNAL}
\author{A.~Bercellie}                     \affiliation{\Rochester}
\author{M.~Betancourt}                    \affiliation{\FNAL}
\author{A.~Bodek}                         \affiliation{\Rochester}
\author{J.~L.~Bonilla}                    \affiliation{\Guanajuato}
\author{A.~Bravar}                        \affiliation{\Geneva}
\author{H.~Budd}                          \affiliation{\Rochester}
\author{G.~Caceres}                       \affiliation{\CBPF}
\author{T.~Cai}                           \affiliation{\Rochester}
\author{M.F.~Carneiro}\thanks{\mateusfcarneiroThanks}  \affiliation{\OregonState}  \affiliation{\CBPF}
\author{G.A.~D\'{i}az~}                   \affiliation{\Rochester}
\author{H.~da~Motta}                      \affiliation{\CBPF}
\author{S.A.~Dytman}                      \affiliation{\Pittsburgh}
\author{J.~Felix}                         \affiliation{\Guanajuato}
\author{L.~Fields}                        \affiliation{\FNAL}\affiliation{\ND}
\author{A.M.~Gago}                        \affiliation{\PUCP}
\author{H.~Gallagher}                     \affiliation{\Tufts}
\author{R.~Gran}                          \affiliation{\UMD}
\author{D.A.~Harris}                      \affiliation{\york}  \affiliation{\FNAL}
\author{S.~Henry}                         \affiliation{\Rochester}
\author{D.~Jena}                          \affiliation{\FNAL}
\author{S.~Jena}                          \affiliation{\Mohali}
\author{J.~Kleykamp}                      \affiliation{\Rochester}
\author{M.~Kordosky}                      \affiliation{\WM}
\author{D.~Last}                          \affiliation{\upenn}
\author{T.~Le}                            \affiliation{\Tufts}  \affiliation{\Rutgers}
\author{A.~Lozano}                        \affiliation{\CBPF}
\author{X.-G.~Lu}                         \affiliation{\oxford}
\author{E.~Maher}                         \affiliation{\MCLA}
\author{S.~Manly}                         \affiliation{\Rochester}
\author{W.A.~Mann}                        \affiliation{\Tufts}
\author{C.~Mauger}                        \affiliation{\upenn}
\author{K.S.~McFarland}                   \affiliation{\Rochester}
\author{B.~Messerly}\thanks{\bamThanks}   \affiliation{\Pittsburgh}
\author{J.~Miller}                        \affiliation{\USM}
\author{J.G.~Morf\'{i}n}                  \affiliation{\FNAL}
\author{D.~Naples}                        \affiliation{\Pittsburgh}
\author{J.K.~Nelson}                      \affiliation{\WM}
\author{C.~Nguyen}                        \affiliation{\Florida}
\author{A.~Norrick}                       \affiliation{\WM}
\author{A.~Olivier}                       \affiliation{\Rochester}
\author{G.N.~Perdue}                      \affiliation{\FNAL}  \affiliation{\Rochester}
\author{M.A.~Ram\'{i}rez}                 \affiliation{\upenn}  \affiliation{\Guanajuato}
\author{H.~Ray}                           \affiliation{\Florida}
\author{H.~Schellman}                     \affiliation{\OregonState}
\author{G.~Silva}                         \affiliation{\CBPF}
\author{C.J.~Solano~Salinas}              \affiliation{\UNI}
\author{H.~Su}                            \affiliation{\Pittsburgh}
\author{M.~Sultana}                       \affiliation{\Rochester}
\author{V.S.~Syrotenko}                   \affiliation{\Tufts}
\author{E.~Valencia}                      \affiliation{\WM}  \affiliation{\Guanajuato}
\author{A.V.~Waldron}                     \affiliation{\ICL}
\author{C.~Wret}                          \affiliation{\Rochester}
\author{B.~Yaeggy}                        \affiliation{\USM}
\author{K.~Yang}                          \affiliation{\oxford}
\author{L.~Zazueta}                       \affiliation{\WM}

\date{\today}
\begin{abstract}
MINERvA presents a new analysis of inclusive charged-current neutrino interactions on a hydrocarbon target. We report single and double-differential cross sections in muon transverse and longitudinal momentum. These measurements are compared to neutrino interaction generator predictions from GENIE, NuWro, GiBUU, and NEUT. In addition, comparisons against models with different treatments of multi-nucleon correlations, nuclear effects, resonant pion production, and deep inelastic scattering are presented. The data recorded corresponds to $10.61\times10^{20}$ protons on target with a peak neutrino energy of approximately 6~GeV. The higher energy and larger statistics of these data extend the kinematic range for model testing beyond previous MINERvA inclusive charged-current measurements. The results are not well modeled by several generator predictions using a variety of input models. 
\end{abstract}

\maketitle
\section{Introduction}
\label{sec:Introduction}
Neutrino oscillation experiments~\cite{Abe:2015awa,Adamson:2016tbq,Acciarri:2015uup,Abe:2011ts} depend on  neutrino interaction models to correct for detector and nuclear effects. Oscillation experiments at a few GeV of mean neutrino energy plan to use an inclusive charged-current (CC) signal to maximize far detector statistical precision. An important component of these measurements is the identification of the resulting lepton. Accurate prediction of the momentum and angular distributions of the lepton are required to correct the measured rate for efficiency and acceptance in both near and far detectors. Models of neutrino interactions are also used as input to neutrino energy reconstruction; mismodeling of lepton energy is \textit{prima facie} evidence that neutrino energy reconstruction will be similarly flawed when using that neutrino interaction model as an input.

Inclusive cross section measurements have been made on a variety of nuclear targets in the past. MicroBooNE~\cite{Adams:2019iqc} and T2K~\cite{Abe:2018uhf} have measured double-differential cross sections as a function of muon momentum and angle on argon and hydrocarbon, but at a lower mean neutrino energy than \minerva~\cite{Aliaga:2013uqz}. NOMAD~\cite{Wu:2007ab}
as well as MINOS~\cite{Adamson:2009ju} and CCFR~\cite{Seligman:1997fe} made measurements as a function of neutrino energy on carbon and iron respectively. \minerva~has made measurements as a function of neutrino energy using the low-$\nu$ method on carbon for both neutrino and anti-neutrino beams~\cite{DeVan:2016rkm,Ren:2017xov} and as a function of muon transverse and longitudinal momentum in the Low Energy (LE) NuMI beam with a neutrino flux peaked at 3~GeV~\cite{Filkins:2020xol}. The result presented here increases the phase space accepted into the multi-GeV regime and as a result expands the range of transverse momentum from 2.5 to 4.5 GeV and longitudinal momentum from 20 to 60 GeV with a $\sim$12 times larger sample size and flux normalization uncertainty of approximately 1/2 the size of the previous result.

\begin{figure}[!h]
    \centering
    \includegraphics[width=0.95\linewidth]{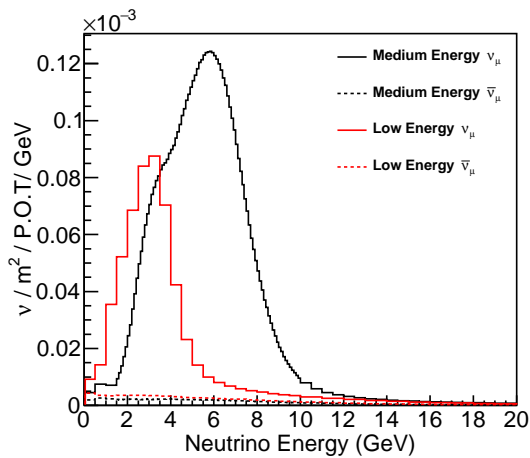}
    \caption{Medium and Low Energy fluxes in the neutrino focused mode at \minerva. In addition to the $\nu_\mu$ flux, the $\bar{\nu_\mu}$ contamination is shown.}
    \label{fig:flux}
\end{figure}

We present here the two-dimensional cross section for the inclusive neutrino scattering as a function of the muon transverse (\pt) and longitudinal momentum (\pz) in the Medium Energy (ME) NuMI beam, which has a neutrino flux peaked near 6~GeV. Figure \ref{fig:flux} compares the Low and Medium Energy fluxes used by \minerva. The muon momentum and angle can be precisely measured. These muon variables are suitable for comparison to exclusive channel measurements and provide a foundation to understand how model predictions combine to form an inclusive cross section prediction. In addition to the two-dimensional cross sections,  one-dimensional projections, limited to the phase space of the double-differential cross section, are also provided. 

Section \ref{sec:Experiment} describes the experimental setup. Section \ref{sec:Simulation} describes the simulation of the neutrino interactions, the modifications made to the interaction model, and the simulation of particle propagation through the detector. The event selection and measurement methods used to extract the differential cross sections are described in Section \ref{sec:extraction}. A description of the sources and determination of systematic uncertainties are presented in Section \ref{sec:SystematicUncertainties}. Section \ref{sec:Results} describes the cross section results while Section \ref{sec:comparisons} provides a set of comparisons to multiple neutrino generator predictions as well as modifications to these predictions. Finally, Section \ref{sec:Conclusion} provides conclusions that can be drawn from these comparisons.

\section{Experiment}
\label{sec:Experiment}
The \minerva~experiment employs a fine-grained tracking detector for recording neutrino interactions produced by the NuMI beamline at Fermilab~\cite{Adamson:2015dkw,Aliaga:2016oaz}. 
Neutrinos are created by directing 120~GeV protons from the Main Injector onto a graphite target. 
The resulting charged pions and kaons are focused by two magnetic horns. 
A neutrino-dominated or anti-neutrino-dominated beam is produced by switching the polarity of the horns.  This analysis uses data from neutrino-dominated beam.

The \minerva~detector~\cite{Aliaga:2013uqz} consists of 120 hexagonal modules that create an active tracking volume preceded by a set of passive nuclear targets. This result includes only those interactions in the active tracking volume with a fiducial mass of 5.48 tons. The active target volume is surrounded by electromagnetic and hadronic calorimeters.

Each tracking module is made of two planes. Each plane is comprised of triangular polystyrene scintillator strips with a 1.7 cm strip-to-strip pitch. To allow for better three-dimensional reconstruction in a high-multiplicity environment, planes are oriented in three different directions, 0$^{\circ}$ and $\pm$ 60$^{\circ}$ relative to the vertical axis of the detector. The downstream and side electromagnetic calorimeter consists of alternating layers of scintillator and 2~mm thick lead planes. The downstream and side hadronic calorimeters consists of alternating scintillator and 2.54~cm thick steel planes.

Multi-anode photomultiplier tubes read out the scintillator strips via wavelength-shifting fibers. The timing resolution measured by thoroughgoing muons is 3.0 ns and sufficient to separate multiple interactions within a single NuMI beam spill. Muons that originate in MINERvA are analyzed by the MINOS near detector~\cite{Michael:2008bc}, a magnetized spectrometer composed of scintillator and iron and located 2~m downstream of the \minerva~detector. The requirement that muons are analyzed in MINOS restricts this analysis to muons with \pz~$ >1.5$~GeV/c and $\theta_\mu<20^\circ$, which means a restricted acceptance for events with $Q^2\stackrel{<}{\sim}\frac{p_{||}^2}{8}$.

This analysis uses data that correspond to $10.61\times 10^{20}$ protons on target (POT), received between September 2013 and February 2017 while the horn polarity was set to focus positively charged particles, creating a beam that is predominantly muon neutrinos. 

\section{Simulation}
\label{sec:Simulation}

A GEANT4-based simulation of the NuMI beamline is used to predict the neutrino flux. To improve the prediction, the simulation is reweighted as a function of pion kinematics to correct for differences between the GEANT4~\cite{Agostinelli:2002hh}\footnote{The MINERvA beam simulation uses GEANT4 version 4.9.2.p3 with the FTFP BERT physics list.} prediction and hadron production measurements of 158~GeV protons on carbon from the NA49 experiment~\cite{Alt:2006fr} and other relevant hadron production measurements. A description of this procedure is found in Ref.~\cite{Aliaga:2016oaz}. In addition, an \textit{in situ} measurement of neutrino scattering off atomic electrons is used, as described in Ref.~\cite{Valencia:2019mkf}, to constrain the flux prediction.

Neutrino interactions are simulated using the GENIE neutrino event generator~\cite{Andreopoulos:2009rq} version 2.12.6. Quasi-elastic (1p1h) interactions are simulated using the Llewellyn-Smith formalism~\cite{LlewellynSmith:1971zm} with the vector form factors modeled using the BBBA05 model~\cite{Bradford:2006yz}. The axial vector form factor uses the dipole form with an axial mass of $M_A=0.99$ GeV/c$^2$. Resonance production is simulated using the Rein-Sehgal model~\cite{Rein:1980wg} with an axial mass of $M_A^{RES}=1.12$ GeV/c$^2$. 
Higher invariant mass interactions are simulated using a leading order model for deep inelastic scattering (DIS) with the Bodek-Yang prescription~\cite{Bodek:2004pc} for the modification at low momentum transfer squared, $Q^2$.

A relativistic Fermi gas model~\cite{Smith:1972xh} is used with an additional Bodek-Ritchie high momentum tail~\cite{Bodek:1981wr} to account for nucleon-nucleon short range correlations. The maximum momentum for Fermi motion is assumed to be $k_F=0.221$ GeV/c.  GENIE models intranuclear rescattering, or final state interactions (FSI), of the produced hadrons using the INTRANUKE-hA package~\cite{Dytman:2007zz}.

To better describe \minerva~data, a variety of modifications to the interaction model are made. To better simulate quasielastic events, the cross section is modified as a function of energy and three momentum transfer based on the random phase approximation (RPA) part of the Valencia model~\cite{Nieves:2004wx,Gran:2017psn} appropriate for a Fermi gas~\cite{Martini:2016eec,Nieves:2017lij}. Multi-nucleon scattering (two-particle two-hole or ``2p2h'') is simulated by the same Valencia model~\cite{Nieves:2011pp,Gran:2013kda,Schwehr:2016pvn}, but the cross section is increased in specific regions of 
energy and three momentum transfer based on fits to \minerva~ data~\cite{Rodrigues:2015hik} in a lower energy beam configuration. Integrated over all phase space, the rate of 2p2h is increased by 50\% over the nominal prediction. Based on fits done in Ref.~\cite{Rodrigues:2016xjj}, we decrease the non-resonant pion production by 43\% and reduce the uncertainty compared to the base GENIE model uncertainties. This modified version of the simulation is referred to as \tune.

The response of the \minerva~detector is simulated using GEANT4~\cite{Agostinelli:2002hh} version 4.9.3.p6 with the QGSP\_BERT physics list. The optical and electronics performance is also simulated. Through-going muons are used to set the absolute energy scale of minimum ionizing energy depositions by requiring the average and RMS of energy deposits match between data and simulation as a function of time. A full description is found in Ref.~\cite{Aliaga:2013uqz}. Measurements using a charged particle test beam~\cite{Aliaga:2015aqe} and a scaled-down version of the \minerva~detector set the absolute energy response to charged hadrons. The effects of accidental activity are simulated by overlaying hits in both \minerva~and MINOS from data corresponding to random beam spills appropriate to the time periods in the simulation. 

\section{Cross Section Extraction}
\label{sec:extraction}
A sample of neutrino charged-current interactions is extracted by requiring the track identified as being from a muon to be matched between \minerva~and MINOS, and to be negatively charged. In addition, the reconstructed interaction vertex must be within a specified fiducial volume. To avoid model dependence introduced by correcting for kinematic regions without acceptance, we only report results for charged-current cross section where the muon angle with respect to the neutrino direction is less than $20^{\circ}$, the muon \pt~is less than 4.5 GeV, and the muon \pz~is between 1.5 GeV and 60 GeV. 

\begin{figure*}
    \centering
        \includegraphics[width=\textwidth] {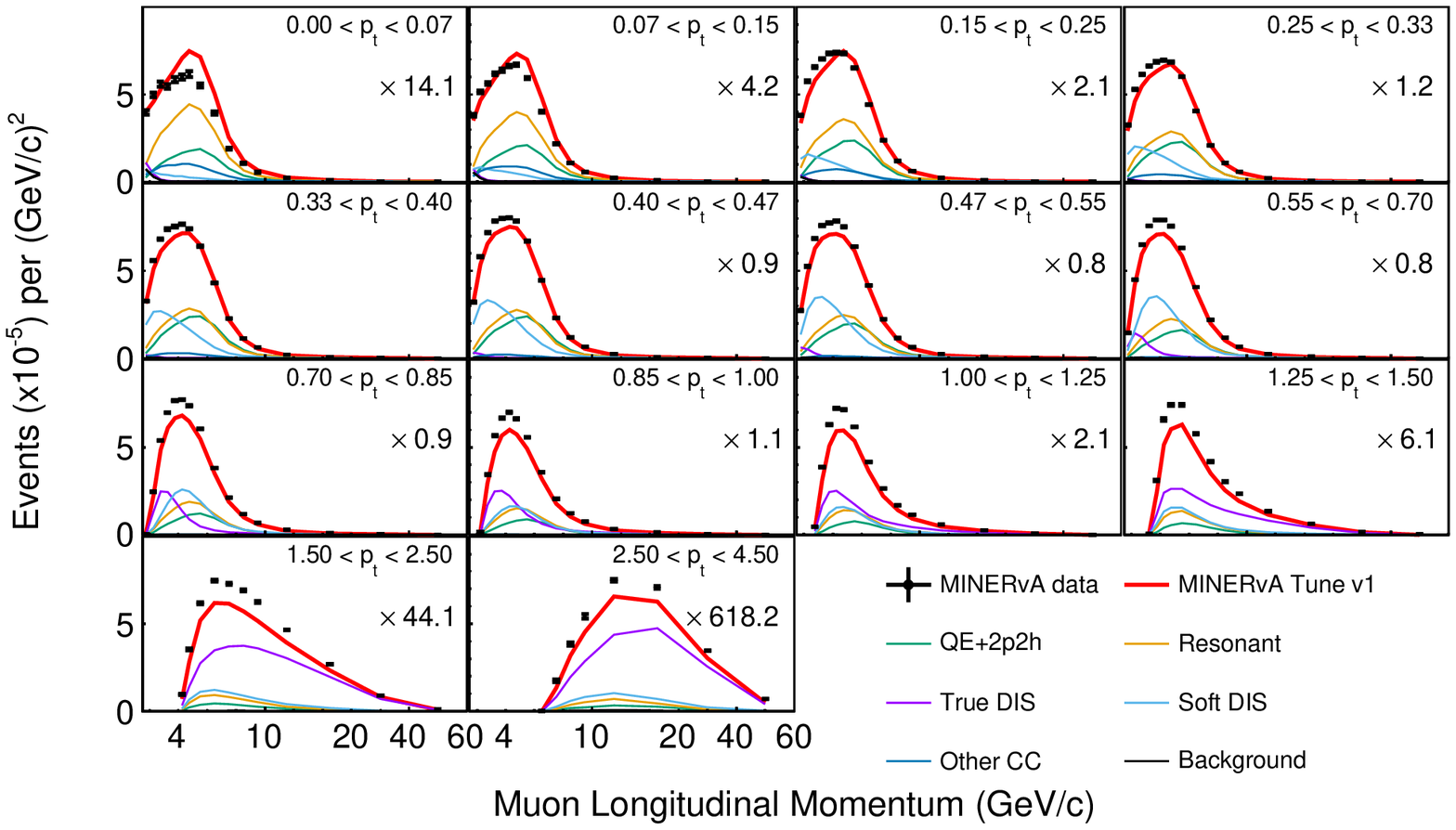}
        \includegraphics[width=\textwidth] {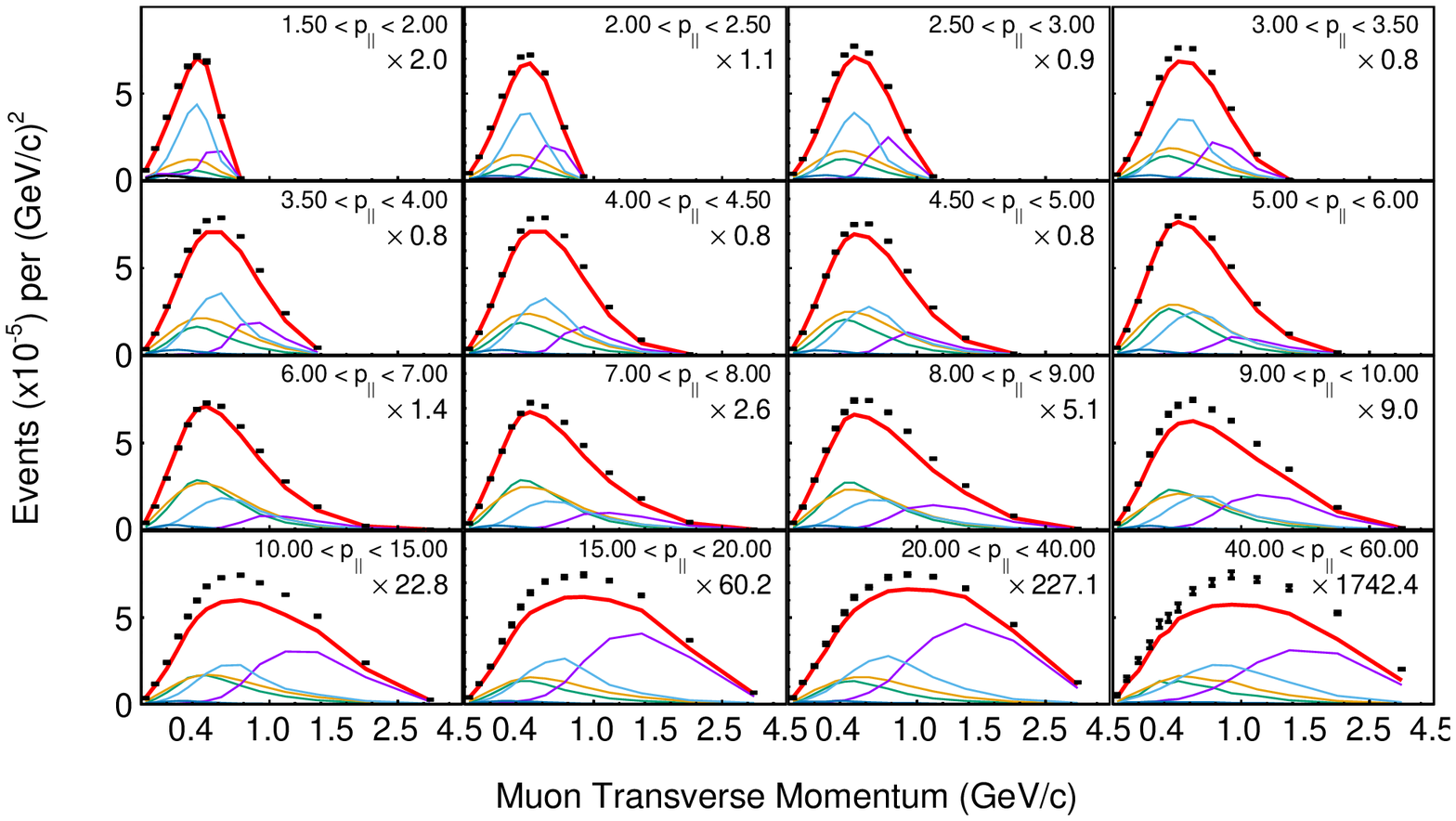}
    \caption{Selected events passing all cuts in data (black points) and simulation (red line) .  Predictions from the simulation, \tune, for various sample components (unstacked), in particular ``Soft DIS", are based on the GENIE generator and defined in Sec. \ref{sec:extraction}. The indicated scale factors are applied to individual panel contents. The x-axis binning reduces the width of the largest \pt~and \pz~ bins for visual compactness. Only statistical uncertainty is shown.}
    \label{fig:selectedSample}
\end{figure*}

Using these criteria, a sample of 4,105,696 interactions was selected. The simulation predicts an average selection efficiency of 64\% in the \pt-\pz~phase space, where the efficiency loss is due to the \minerva-MINOS geometric acceptance. After all selection cuts, the sample in muon transverse and longitudinal momentum space is shown in Fig. \ref{fig:selectedSample}, decomposed into predicted components. Events are labeled by categories within GENIE except for events given a DIS label. To explore how contributions from DIS events rely on the validity of the neutrino-quark scattering model with different ``depth" of the inelasticity,
DIS is divided into two categories. ``True DIS" events are those events where the invariant mass of the hadronic system, $W$, is greater than 2.0 GeV/$c^2$ and $Q^{2}$ greater than 1.0 GeV$^{2}$/c$^{4}$. ``Soft DIS" represents the remainder of the GENIE DIS events. While the cut defining the ‘True DIS’ regime is defined by $Q^2$ and hadronic invariant mass $W$ values, and could therefore be used for model comparisons, the ‘Soft DIS’ definition is not.    The modeling of inelastic events below the ‘True DIS’ region can vary widely across generators, in terms of the kinematic coverage of the resonance model and handling of non-resonant contributions in the resonance region \cite{SajjadAthar:2020nvy}.    No two generators handle these aspects in exactly the same way, so the ‘Soft DIS’ label here is relevant only to GENIE simulations. A final category is ``other CC", which contains CC events not belonging to the other categories, such as coherent charged pion production. 
The background category contains charged-current events from other neutrino flavors and anti-muon type neutrinos as well as neutral-current interactions. A total of 8655 (0.2\%) background events are predicted. Backgrounds at \pz~less than 2.5 GeV and \pt~less than 0.4~GeV are primarily from neutral-current interactions where a pion was reconstructed as a muon in MINOS. These backgrounds have a maximum contribution of 10 percent of the predicted event rate at these low \pz~and \pt, and are typically much smaller. Backgrounds at high \pt~are mostly anti-neutrino contamination due to muon charge misidentification which accounts for about one percent of the sample in the highest \pt~bin.

\begin{figure}[!h]
    \centering
    \includegraphics[width=0.95\linewidth] {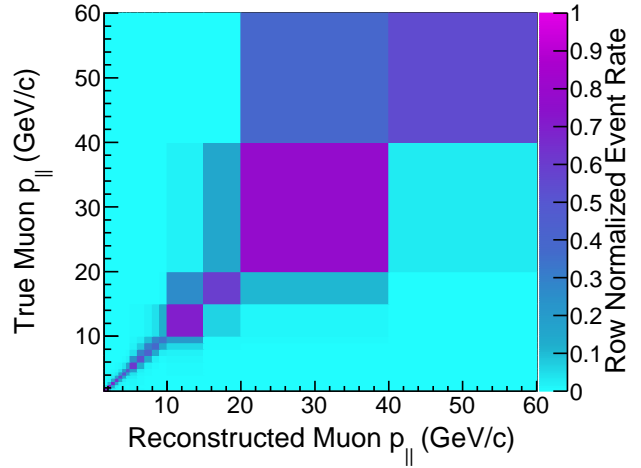}
    \caption{Event migration between simulated and reconstructed \pz~bins projected over all \pt.}
    \label{fig:pzmig}
\end{figure}

\begin{figure}[!h]
    \centering
    \includegraphics[width=0.95\linewidth] {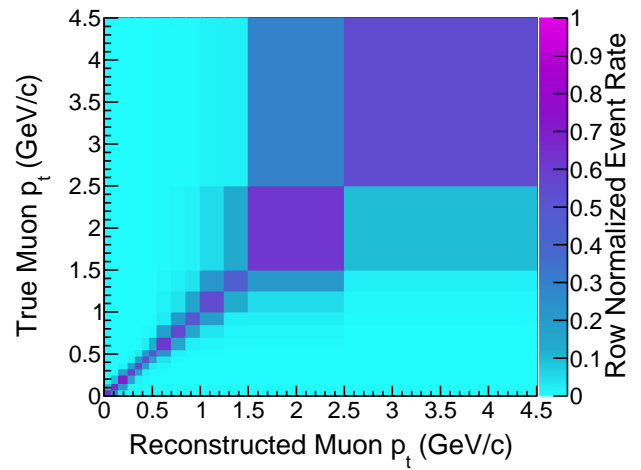}
    \caption{Event migration between simulated and reconstructed \pt~bins projected over all \pz.}
    \label{fig:ptmig}
\end{figure}

The predicted background contributions are subtracted from the sample. Detector resolution effects (see Figs. \ref{fig:ptmig} and \ref{fig:pzmig}) are then removed using the D'Agostini unfolding method~\cite{D'Agostini:1994zf,DAgostini:2010xxxxx}, via the implementation in RooUnfold~\cite{Adye:2011gm}. To understand the necessary regularization strength, 10 different model predictions as pseudodata
were unfolded using the \tune. These fake data were derived by reweighting the 2p2h strength, QE RPA, non-resonsant pion production reweight, resonant pion production. Many of the models used appear in Table \ref{tab:DDModelComp}. The unfolded models were then compared to their true distributions via a $\chi^{2}$ test taking full consideration of correlations. The optimal number of iterations was determined when the $\chi^{2}$ approached one per degree of freedom and was not changing as a function of the number of iterations. Only statistical uncertainties were considered in determining the number of iterations. In all variations the required number of iterations was no more than 10. In addition, a fit to the data in reconstructed \pt-\pz~was performed, and the \tune~prediction reweighted to the data as an additional fake data sample.A fit is performed as a function of \pt-\pz~to provide a reweight value forcing the \tune~to better agree with the data. This reweighted prediction is used as another fake data sample. Reweighting is done in true kinematic quantities, although the fit is done with reconstructed quantities, and propagated through the Monte Carlo detector response prediction. Based on these studies, the data was unfolded using 10 iterations.

\begin{figure}[!h]
    \centering
    \includegraphics[width=0.95\linewidth] {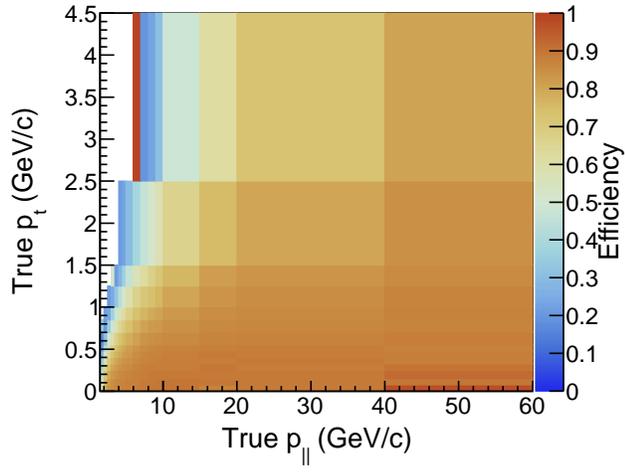}
    \caption{Selection efficiency as a function of \pt~and \pz.}
    \label{fig:selectionEfficiency}
\end{figure}

Finally, the sample is corrected for efficiency and acceptance. The selection efficiency is shown in Fig.~\ref{fig:selectionEfficiency}. The large efficiency in the 6-7 GeV \pz~and highest \pt~bin is due to a fractionally large sample of muons with generated angle greater than 20 degrees passing event selection and appearing in this bin.
 
The efficiency-corrected distribution is then divided by the integral of the flux with neutrino energies between 0 and 100~GeV averaged over the fiducial volume, which is $6.32\times10^{-8}\pm 3.9\%$ per cm$^2$ per proton on target, and the number of nucleons in the fiducial volume, $3.23\times10^{30}\pm 1.4\%$, with a mass fraction of 88.51\% carbon, 8.18\% hydrogen, 2.5\% oxygen, 0.47\% titanium, 0.2\% chlorine, 0.07\% aluminum, and 0.07\% silicon.

\newpage
\clearpage
\section{Systematic Uncertainties}
\label{sec:SystematicUncertainties}

Systematic uncertainties in this analysis fall under three different categories:  flux, detector response, and neutrino interaction model uncertainties.  
The uncertainties from individual sources are evaluated by re-extracting the cross section using modified simulations.  The size of each modification is related to the uncertainty in each source. Flux uncertainty, a typical leading uncertainty in neutrino cross section measurements, is below 4\% for almost all the phase space because of the flux constraint that comes from a measurement of neutrino-electron scattering in the same beam~\cite{Valencia:2019mkf}. The normalization uncertainty of 1.4\% corresponds to the  uncertainty in the number of target nucleons and is based on material assays and weight measurements of production-quality scintillator planes.  

Uncertainty in the detector response to hadrons is evaluated using shifts determined by \textit{in situ} measurements of a smaller version of the detector in a test beam~\cite{Aliaga:2015aqe}. Uncertainties in inelastic interaction cross sections for particles in the detector material are independently varied based on data-Monte Carlo differences between GEANT particle cross sections and world data on neutrons \cite{Abfalterer:2001gw,Schimmerling:1973bb,Voss:1956,Zanelli:1981zz}, pions \cite{Ashery:1981tq,Allardyce:1973ce,Wilkin:1973xd,Clough:1974qt}, and protons \cite{Menet:1971zz,Dicello:1970mx,McGill:1974zz}. 

Muon reconstruction uncertainty is dominated by the muon energy scale uncertainty, which is constrained by a fit to the reconstructed neutrino energy distribution for low recoil neutrino charged-current events, whose cross section is known to be flat as a function of neutrino energy. In the low recoil fit procedure, we include the model uncertainties that are used in this and other MINERvA results, including uncertainty on the 2p2h process informed by MINERvA data\cite{Rodrigues:2015hik}. Because the low recoil cut is at 800 MeV, this results in a small uncertainty in the fit due to cross section modeling\cite{Bashyal:stuffhere}.The resulting uncertainty on the muon energy scale is 1\%. The fit causes the flux and muon energy scale uncertainties to be correlated and that correlation is propagated to the final result. Uncertainty in the matching efficiency is from imperfect modeling of the efficiency loss from accidental activity in the MINOS near detector when matching muon tracks from MINERvA to MINOS.  This last efficiency is also determined by a data-simulation comparison as a function of instantaneous neutrino beam intensity.  

Interaction model uncertainties are evaluated using the standard GENIE reweighting infrastructure\cite{Andreopoulos:2009rq,andreopoulos2015genie}. Because this is an inclusive analysis with very low backgrounds and few selection cuts, model uncertainties are never the dominant uncertainty in any \pt-\pz~bin.  These  uncertainties are most significant at the highest \pt~bins where the geometric acceptance changes dramatically and at low \pt~bins where the backgrounds are the largest. The reconstruction of the muon is largely unaffected by the hadronic shower. The efficiency of the selection as a function of \pt-\pz-hadronic energy was evaluated and found to be flat as a function of hadronic energy.

\begin{figure}
    \centering
    \includegraphics[width=0.95\linewidth] {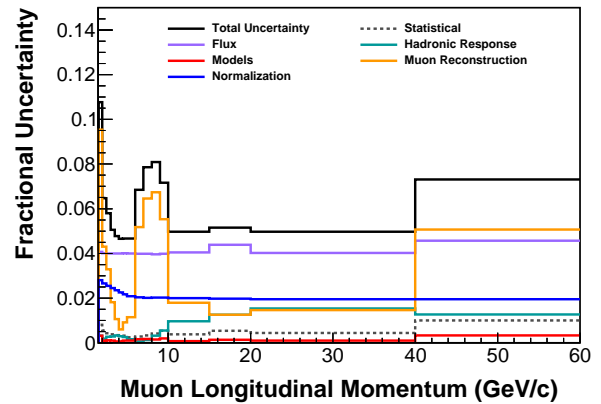}
    \caption{Fractional uncertainty of the cross section when projected on the \pz~axis.}
    \label{fig:pzsys}
\end{figure}

\begin{figure}
    \centering
    \includegraphics[width=0.95\linewidth] {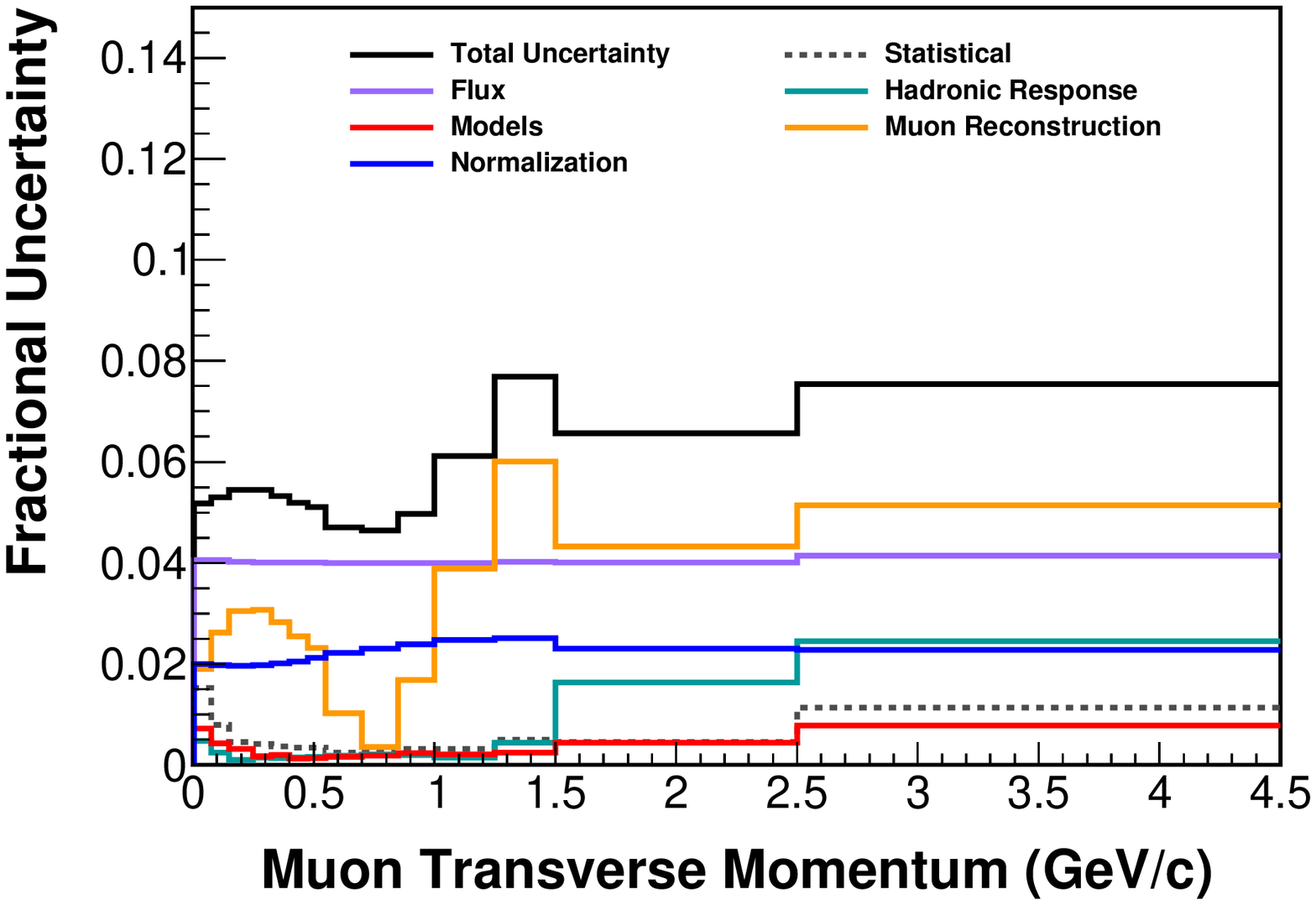}
    \caption{Fractional uncertainty of the cross section when projected on the \pt~axis.}
    \label{fig:ptsys}
\end{figure}

\begin{figure*}[p]
    \centering
    \includegraphics[width=\textwidth] {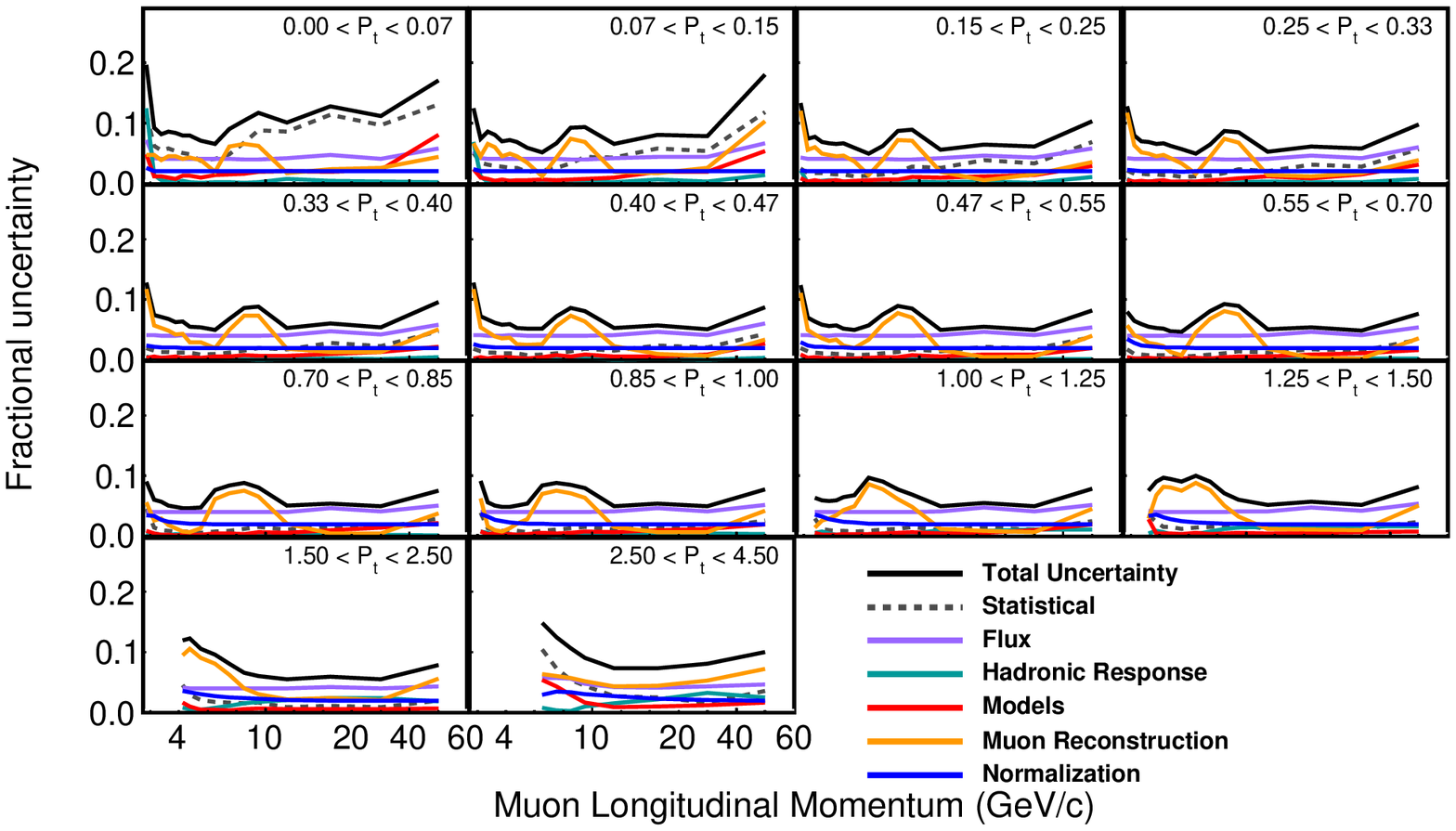}
    \includegraphics[width=\textwidth] {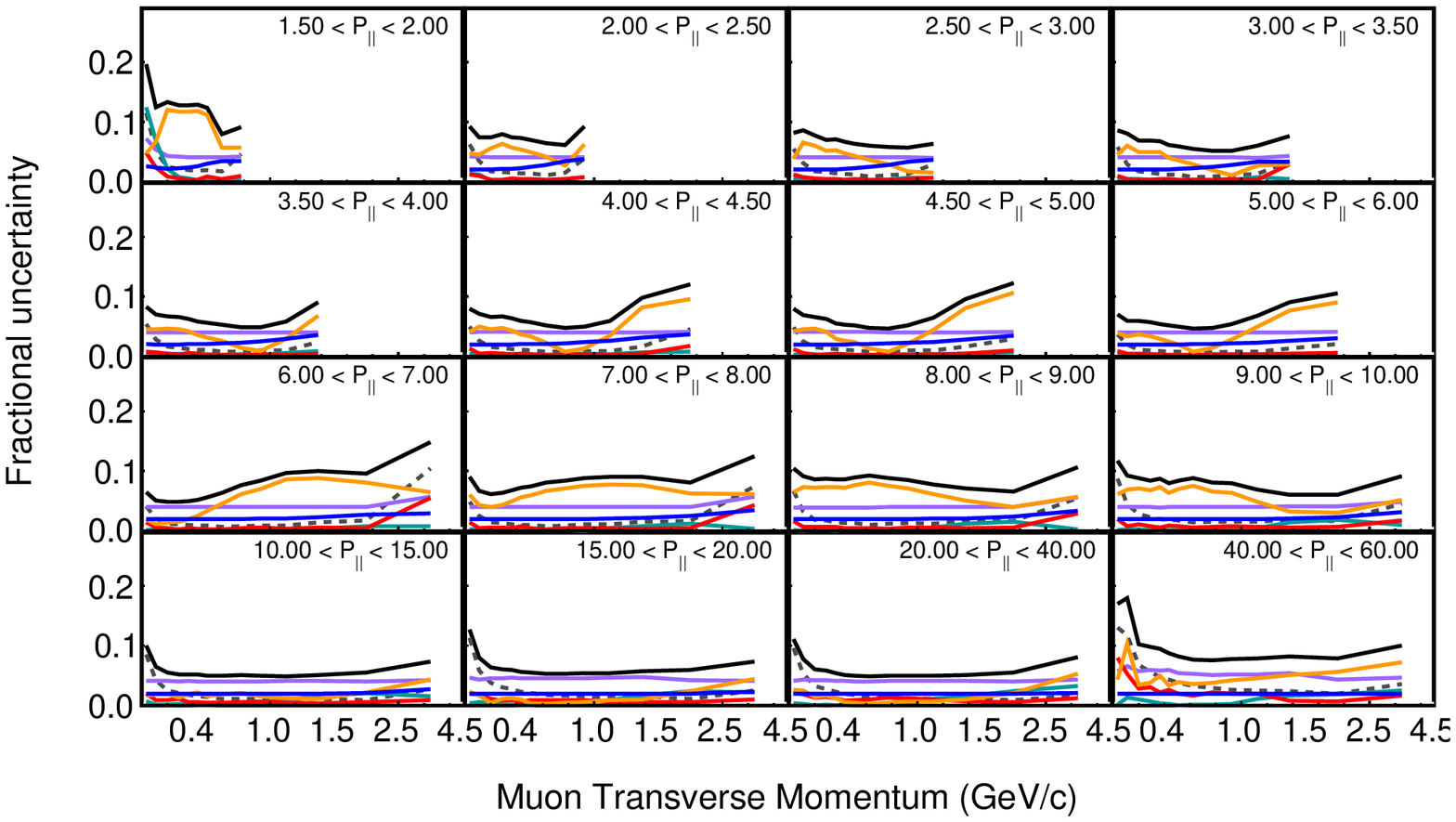}
    \caption{Fractional uncertainties of the two-dimensional cross section as a function of \pt~and \pz.}
    \label{fig:2Dsys}
\end{figure*}

The fractional uncertainties in the one-dimensional projections are shown in Figs.~\ref{fig:pzsys} and \ref{fig:ptsys}. The fractional uncertainties in the two-dimensional result are shown in Fig.~\ref{fig:2Dsys}.  The dominant uncertainties are the muon energy scale uncertainty and the flux normalization. The muon energy scale uncertainty has the largest effect on the cross section measurement at the rising and falling edges of the peak of the muon momentum spectrum, where the slope between bins is largest. The muon momentum peaks at approximately \pz~= 5 GeV and \pt~= 0.6 GeV. It should be noted there are particular regions where the pion re-interaction probability uncertainty, grouped in the hadronic response systematic, is large. This is due to interactions in which the reconstructed muon was not the primary muon, but was instead due to a high energy pion. This is also how a population of neutral current interactions populate the lowest \pz~bins at low \pt.


\newpage
\clearpage

\section{Results}
\label{sec:Results}
\begin{figure}
    \centering
    \includegraphics[width=0.95\linewidth]{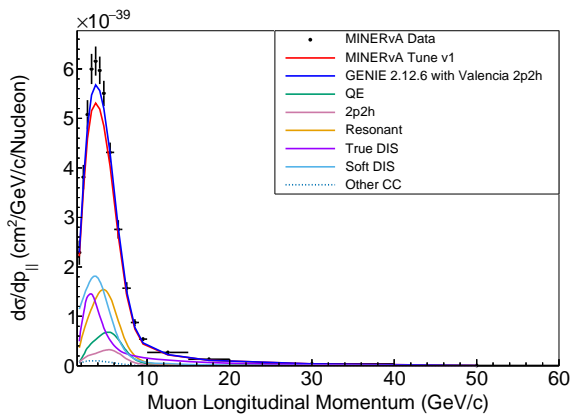}
    \caption{Cross section projected onto the \pz~axis showing as predicted by \tune~where the different contributions, in particular "Soft DIS", are based on the GENIE generator and are defined in Section IV. The GENIE 2.12.6 with Valencia 2p2h prediction is also shown. Total uncertainty is shown.}
    \label{fig:1Dpzxsec}
\end{figure}

\begin{figure}
    \centering
    \includegraphics[width=0.95\linewidth]{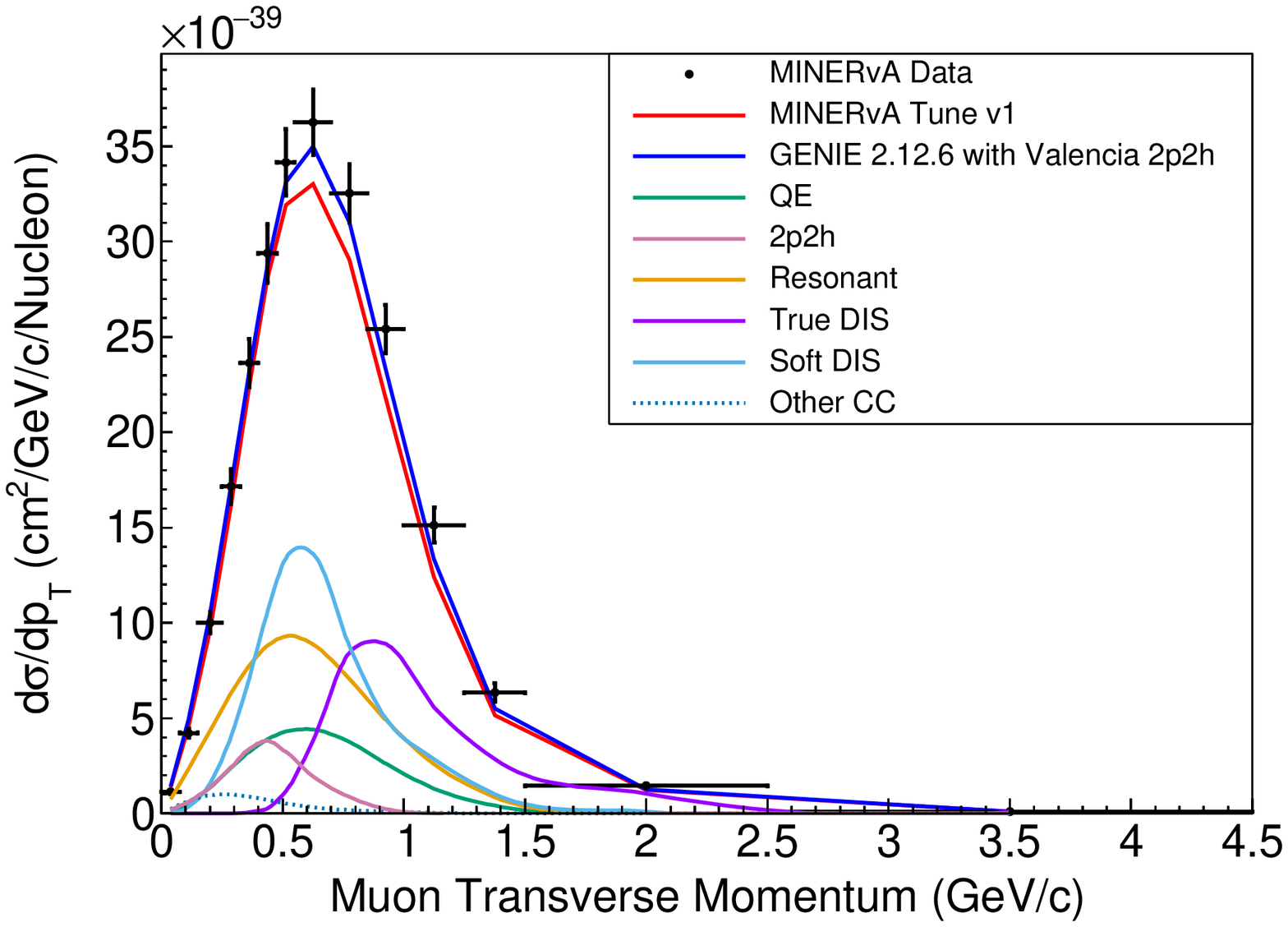}
    \caption{Cross section projected onto the \pt~axis showing contributions as predicted by \tune~where the different contributions, in particular "Soft DIS", are based on the GENIE generator and are defined in Section IV. The GENIE 2.12.6 with Valencia 2p2h prediction is also shown. Total uncertainty is shown. }
    \label{fig:1Dptxsec}
\end{figure}

\begin{figure}
    \centering
    \includegraphics[width=0.95\linewidth]{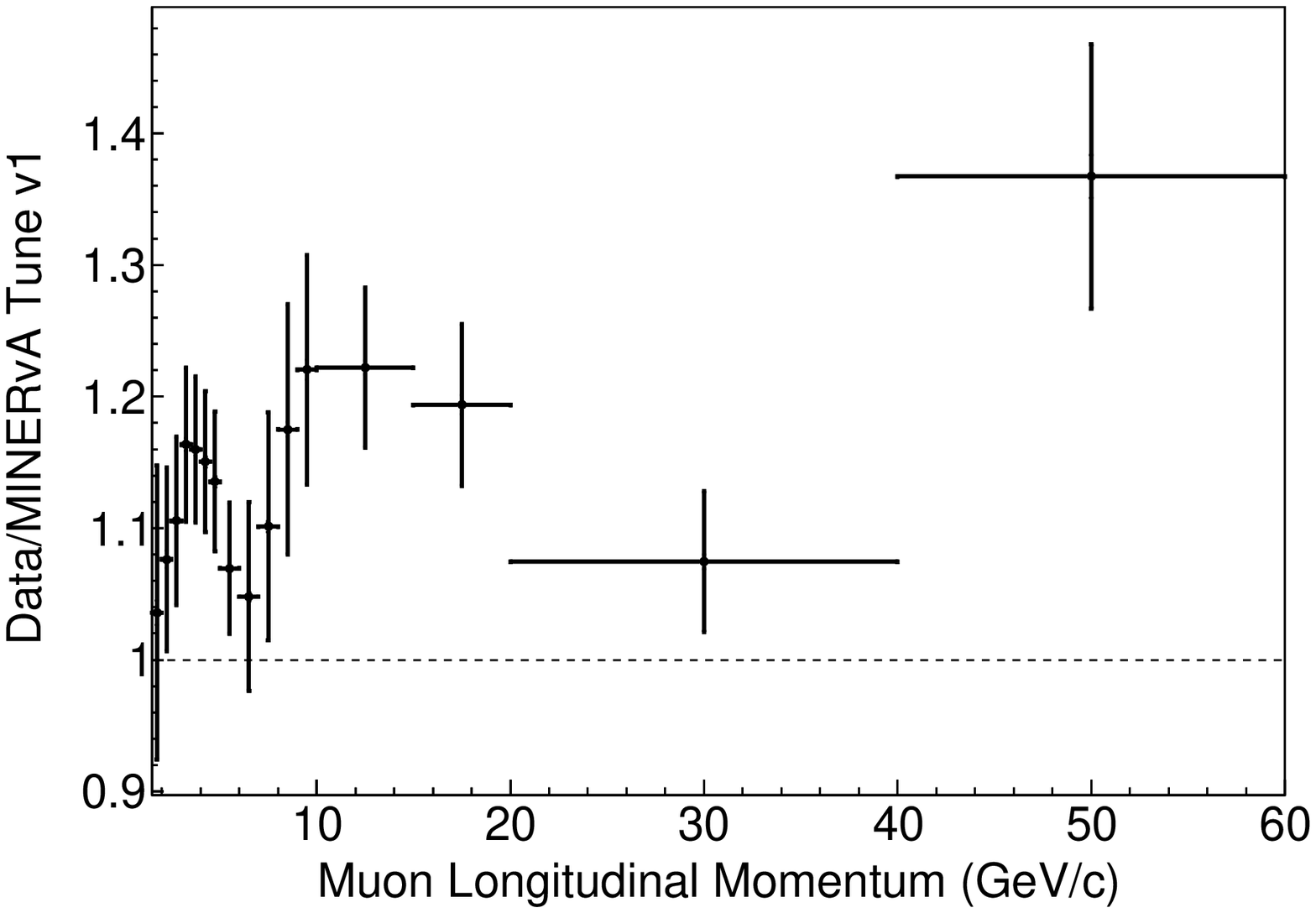}
    \caption{Ratio of measured to \tune~as projected onto the \pz~axis. Total uncertainty is shown.}
    \label{fig:1Dpzratio}
\end{figure}

\begin{figure}
    \centering
    \includegraphics[width=0.95\linewidth]{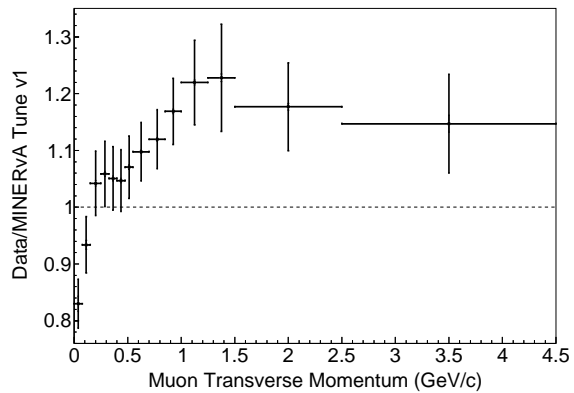}
    \caption{Ratio of measured to \tune~as projected onto the \pt~axis. Total uncertainty is shown.}
    \label{fig:1Dptratio}
\end{figure}

\begin{figure*}
    \centering
    \includegraphics[width=\textwidth] {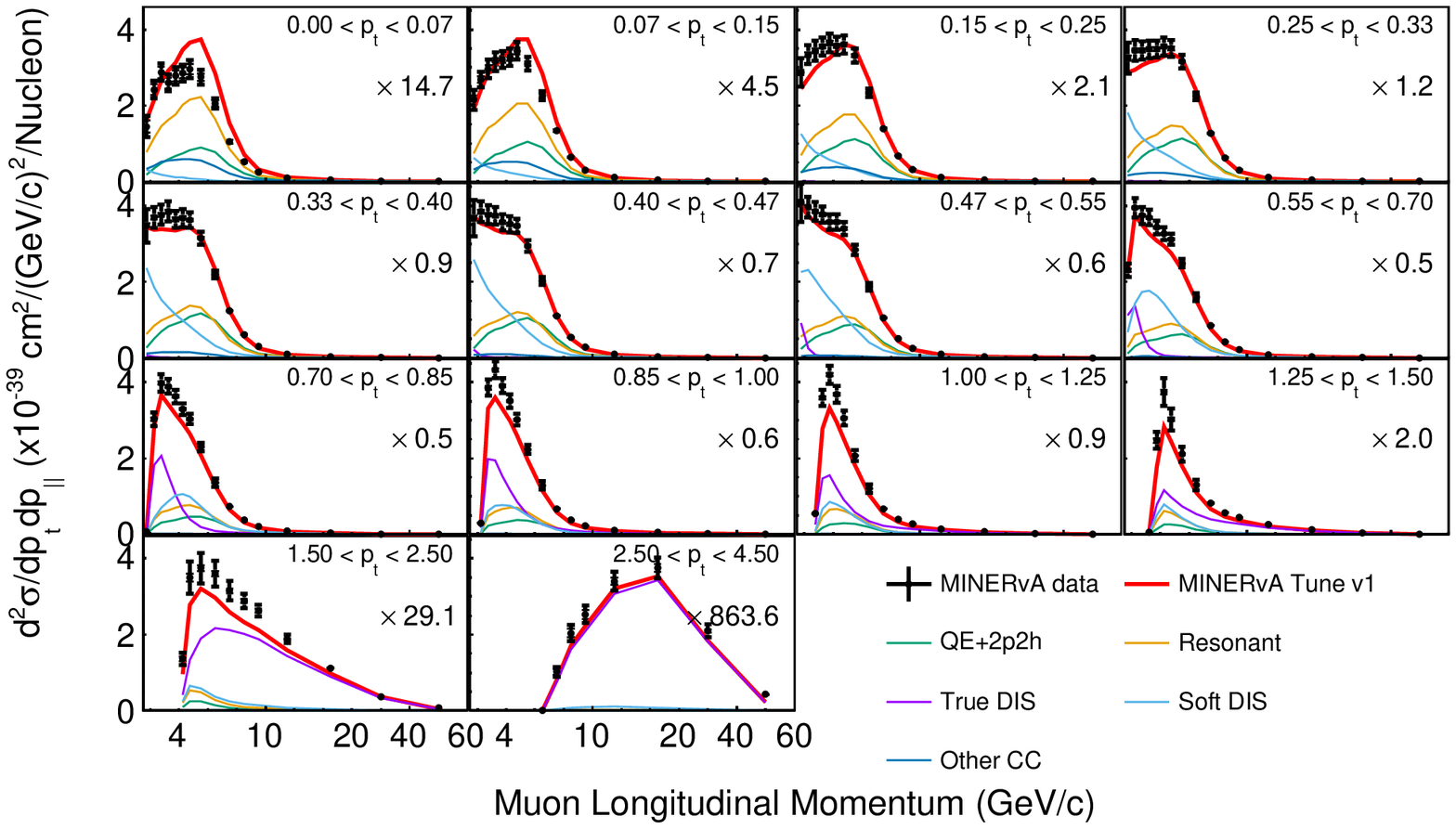}
    \includegraphics[width=\textwidth] {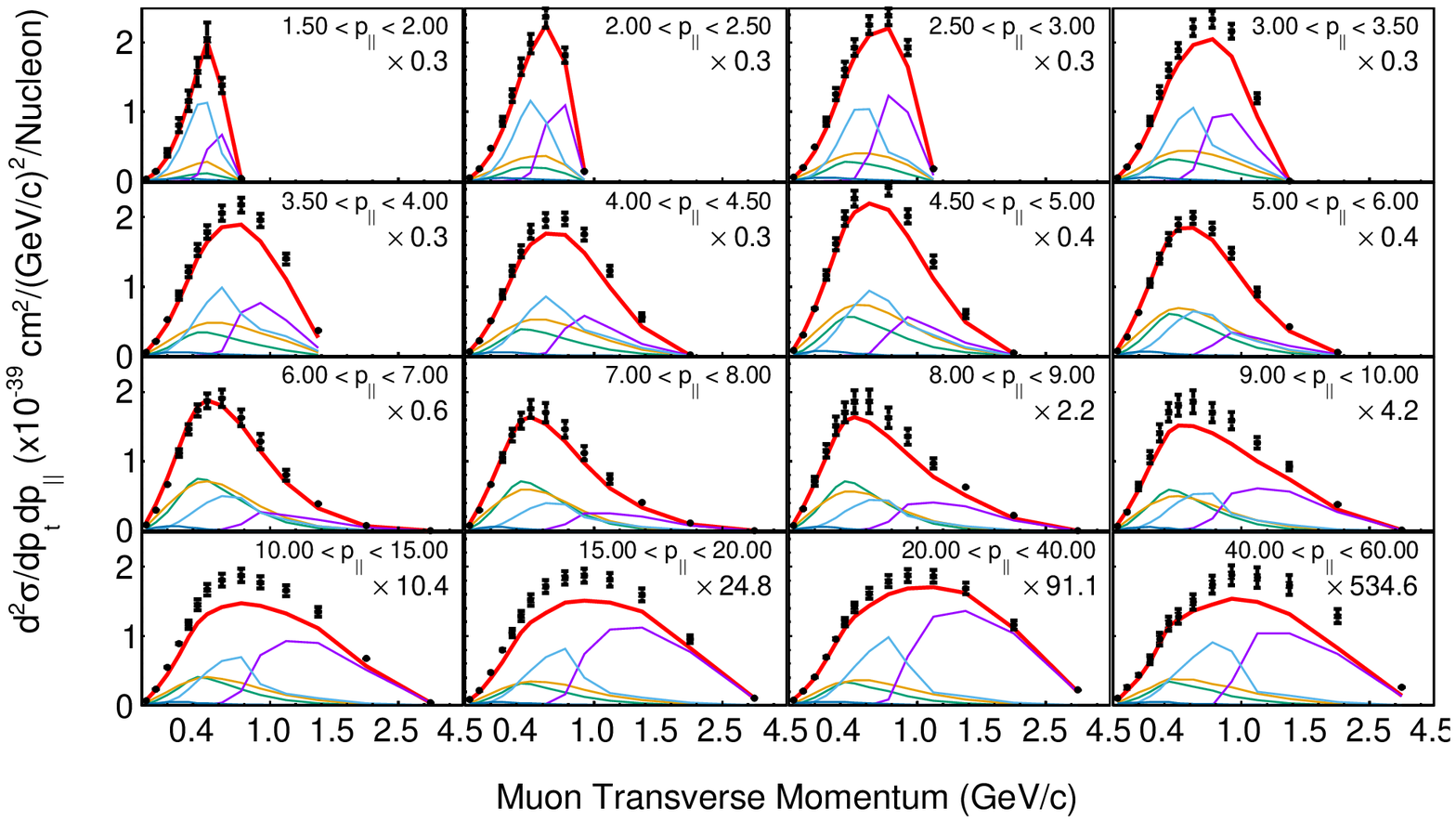}
    \caption{Extracted cross section compared to \tune. Predictions represent \tune, with various sample components (unstacked), in particular ``Soft DIS", based on the GENIE generator and defined in Sec. \ref{sec:extraction}. The indicated scale factors are applied to individual panel contents. The x-axis binning reduces the width of the largest \pt~and \pz~ bins for visual compactness. Inner (outer) ticks denote statistical (total) uncertainty.}
    \label{fig:2Dxsec}
\end{figure*}

Three results are presented: two single-differential cross sections and a double-differential cross section using the \pt~and \pz~of the muon. The single-differential cross sections are shown in Figs.~\ref{fig:1Dpzxsec} and~\ref{fig:1Dptxsec}. The ratio of data to \tune~for the single-differential results is shown in Figs.~\ref{fig:1Dpzratio} and \ref{fig:1Dptratio}. The double-differential cross section as a function of \pt~and \pz~is shown in Fig.~\ref{fig:2Dxsec}.


The single-differential cross sections are derived from the two-dimensional result which means the additional phase space restrictions of the two-dimensional spaces are incorporated. The cross section versus \pt~includes a restriction of 1.5~$\leq$~\pz~$\leq$ 60~GeV/c, while the cross section versus \pz~includes a restriction of \pt~$\leq$ 4.5~GeV/c.

The difference between data and \tune,~shown in Fig. \ref{fig:1Dpzratio}, is due to a mismodeling of the cross section as function of muon kinematics combined with the angular acceptance and muon energy scale. A study was performed by correcting the prediction to match the data in the \pt~projection. The \pz~prediction after this correction was compared to the data and was found to be consistent in both normalization and shape within the muon energy scale uncertainty.

\begin{figure*}
    \centering
    \includegraphics[width=\textwidth] {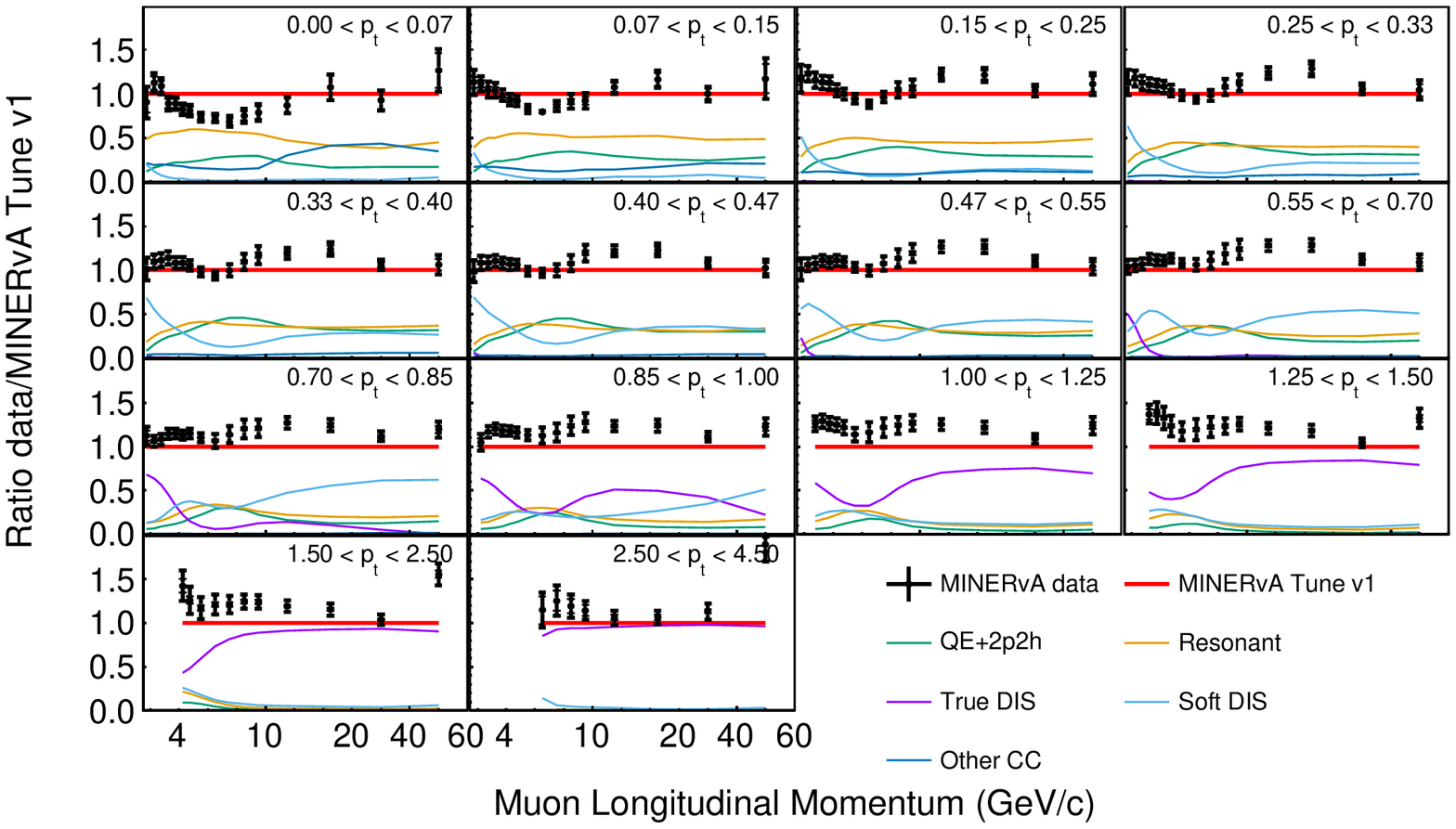}
    \includegraphics[width=\textwidth] {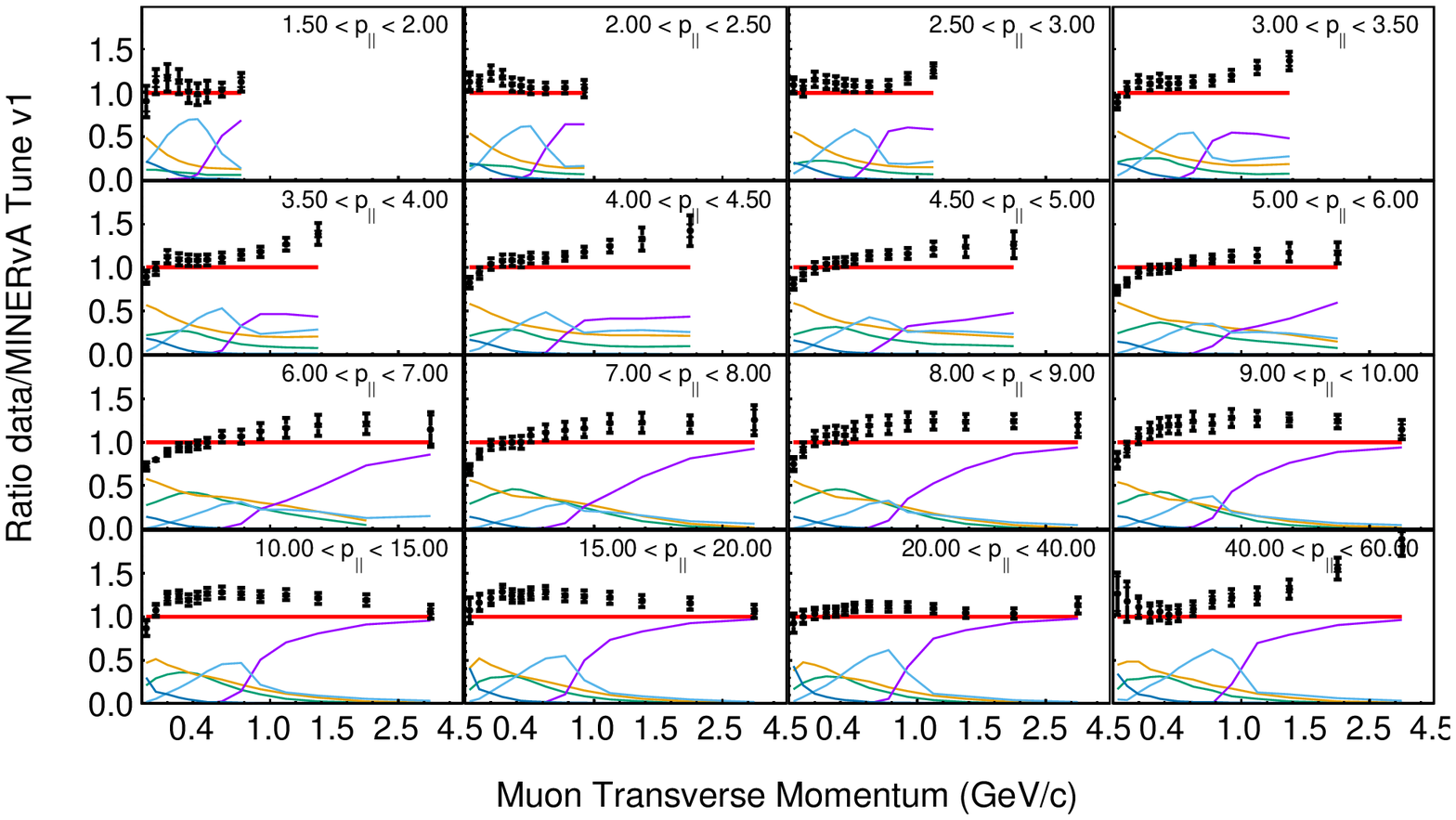}
\caption{Ratio of the extracted cross section to \tune. The various sample components, in particular ``Soft DIS", are based on the GENIE generator and defined in Sec. \ref{sec:extraction} and represent the fractional contribution to the overall prediction. Inner (outer) ticks denote statistical (total) uncertainty.}
\label{fig:2Dxsecratio}
\end{figure*}

Figure \ref{fig:2Dxsecratio} shows the ratio of data to simulation. For \pz~between 3 and 15~GeV and low values of \pt, the cross section is overpredicted. In this region the dominant process is resonant pion production which has been previously measured by \minerva~\cite{Eberly:2014mra, Aliaga:2015wva, McGivern:2016bwh, Altinok:2017xua, Le:2019jfy,Coplowe:2020yea}, MiniBooNE~\cite{AguilarArevalo:2010bm}, and T2K~\cite{Abe:2019arf}. Many of these measurements, including a MINOS measurement~\cite{Adamson:2014pgc}, indicate the need to reduce the predicted cross section at low Q$^2$ which corresponds here to regions of low \pt. A set of comparisons against a variety of resonant pion production model modifications is shown in Sec.~\ref{sec:comparisons}. At \pt~$>$~0.85 GeV/c the Monte Carlo prediction consistently underpredicts the data by 10-25\%. This high \pt~region is dominated by the ``True DIS" process for \pz~$>6$ GeV/c. Poorly understood neutrino DIS nuclear effects could contribute to the underprediction in this region of kinematics.


\section{Comparisons}
\label{sec:comparisons}
In this section the extracted cross sections are compared to a variety of predictions. Four different groups of models and the data are presented as a ratio with respect to \tune.  Each group investigates a different aspect of the \tune. No model combination completely describes the data. We also compare to a group of predictions from different neutrino event generators. 

\begin{figure*}
    \centering
    \includegraphics[width=\textwidth] {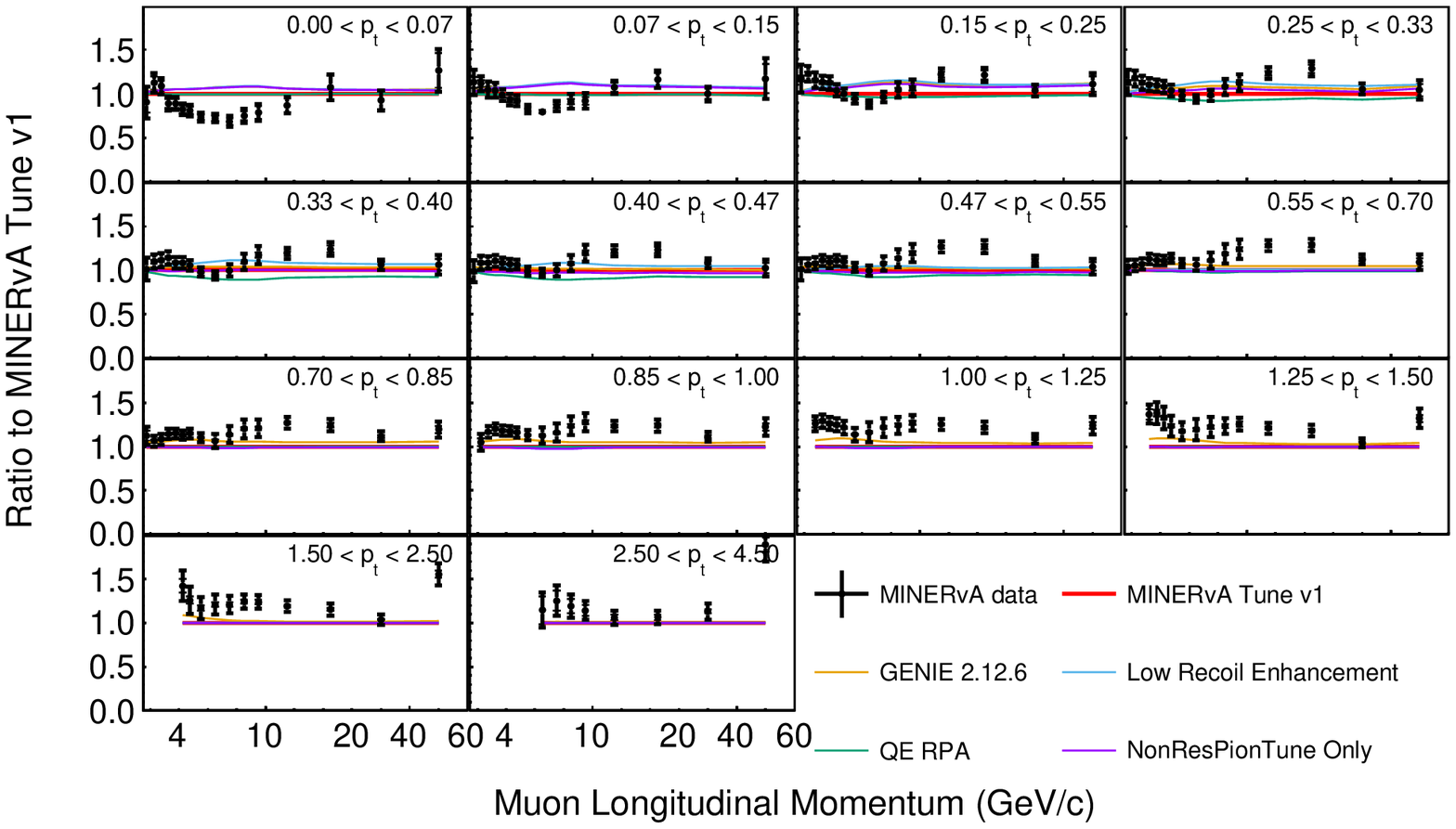}
    \includegraphics[width=\textwidth] {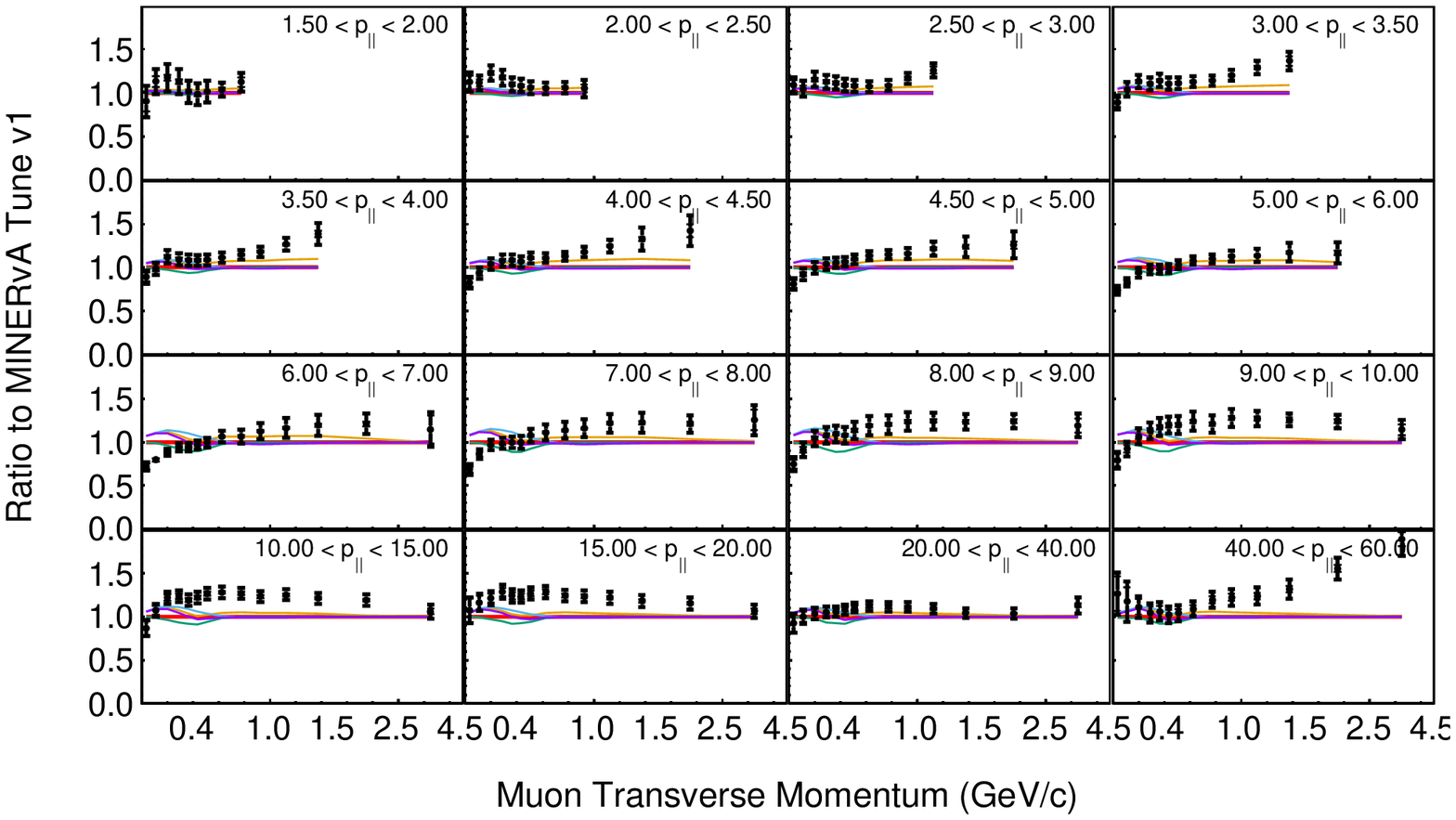}
    \caption{Extracted cross section and four predictions based on specific modifications to components of the \tune~displayed as a ratio to \tune. Inner (outer) ticks denote statistical (total) uncertainty.}
    \label{fig:set0comp}
\end{figure*}

The first of these groups of comparisons, see Fig. \ref{fig:set0comp}, considers alterations to \tune~by adding or subtracting selected changes that were made to the original default GENIE prediction. The cases plotted are:
\begin{enumerate}[label=(\alph*),noitemsep]
    \item GENIE v2.12.6 with Valencia 2p2h \cite{Nieves:2011yp, Sobczyk:2012ms, Gran:2013kda, Schwehr:2016pvn}, (GENIE 2.12.6);
    \item Case a) but with a reduction in the non-resonant pion production of 43\% \cite{Rodrigues:2016xjj}, (NonResPionTune Only);
    \item Case b) but with RPA from the Valencia model applied to quasielastic events, (QE RPA);
    \item Case b) but with the  application of the empirical enhancement of 2p2h production as described in Sec. \ref{sec:Simulation}, (Low Recoil Enhancement);
    \item The full set of corrections, (\tune).
\end{enumerate}
 
\begin{figure*}
    \centering
    \includegraphics[width=\textwidth] {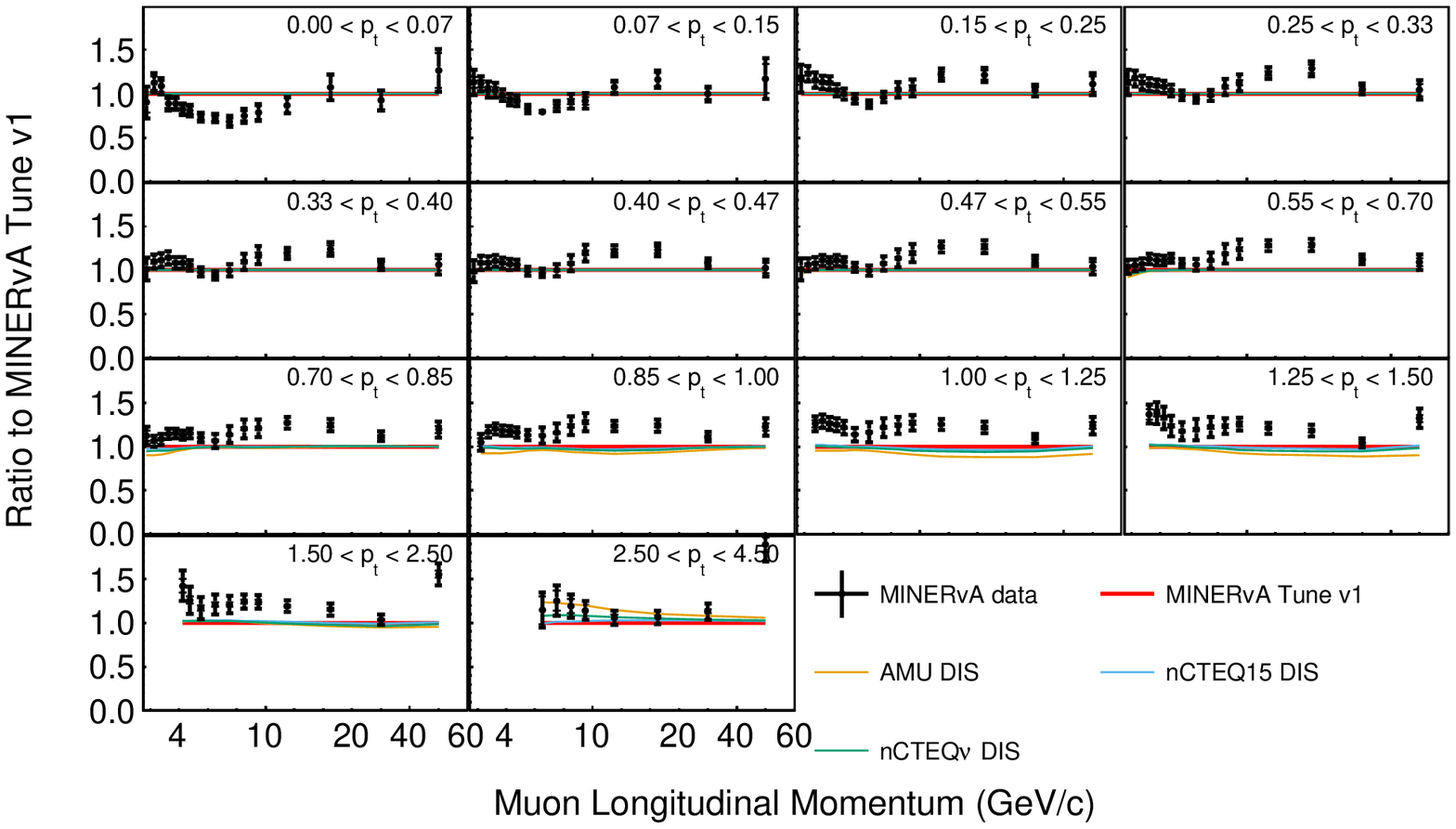}
    \includegraphics[width=\textwidth] {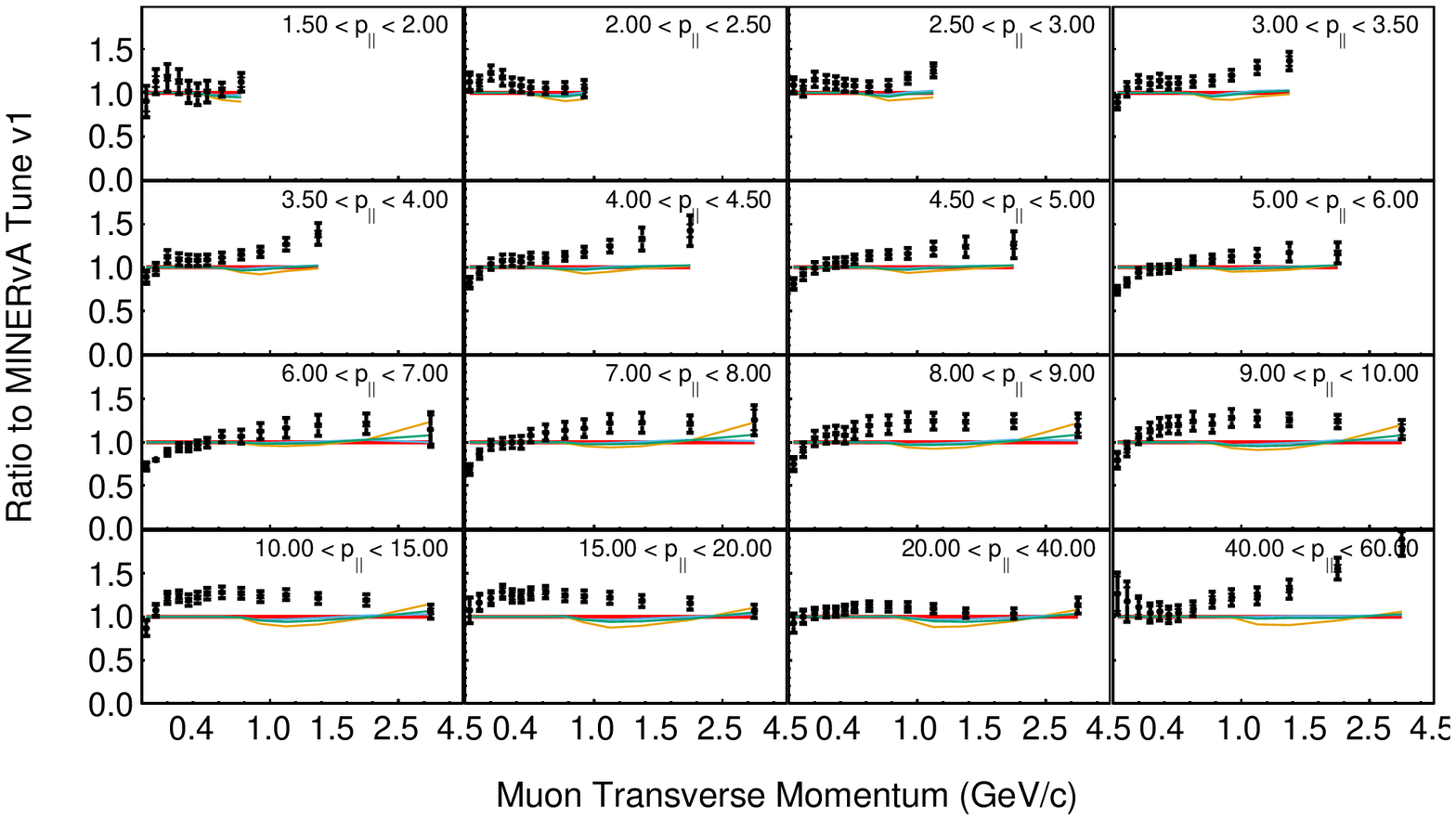}
    \caption{The extracted cross section and predictions that modify the DIS models displayed as a ratio to \tune. Inner (outer) ticks denote statistical (total) uncertainty.}
    \label{fig:set1comp}
\end{figure*}

 The second group of comparisons, see Fig. \ref{fig:set1comp}, use the \tune~as the baseline prediction with modifications to the DIS model for $W\ge$~2.0 GeV/$c^2$ and $Q^{2}\ge$~1.0 GeV$^{2}$/c$^{4}$. The modifications plotted are: 
 \begin{enumerate}[label=(\alph*),noitemsep]
     \item \tune, which uses the LHAPDF5 parton distributions functions (PDFs), (\tune);
     \item Case a) but with the nCTEQ15 PDFs, which are determined from charged lepton-nucleus scattering \cite{Kovarik:2015cma}, rather than the LHAPDF5 PDFs, (nCTEQ15 DIS);
     \item Case a) but with the nCTEQ15$\nu$ pfds, which are determined from neutrino-nucleus scattering \cite{Schienbein:2007fs}, rather than the LHAPDF5 PDFs, (nCETQ$\nu$ DIS);
     \item Case a) but employing the microscopic model developed at Aligarh Muslim University \cite{Haider:2016zrk}, (AMU DIS).
 \end{enumerate}
 
\begin{figure*}
    \centering
    \includegraphics[width=\textwidth] {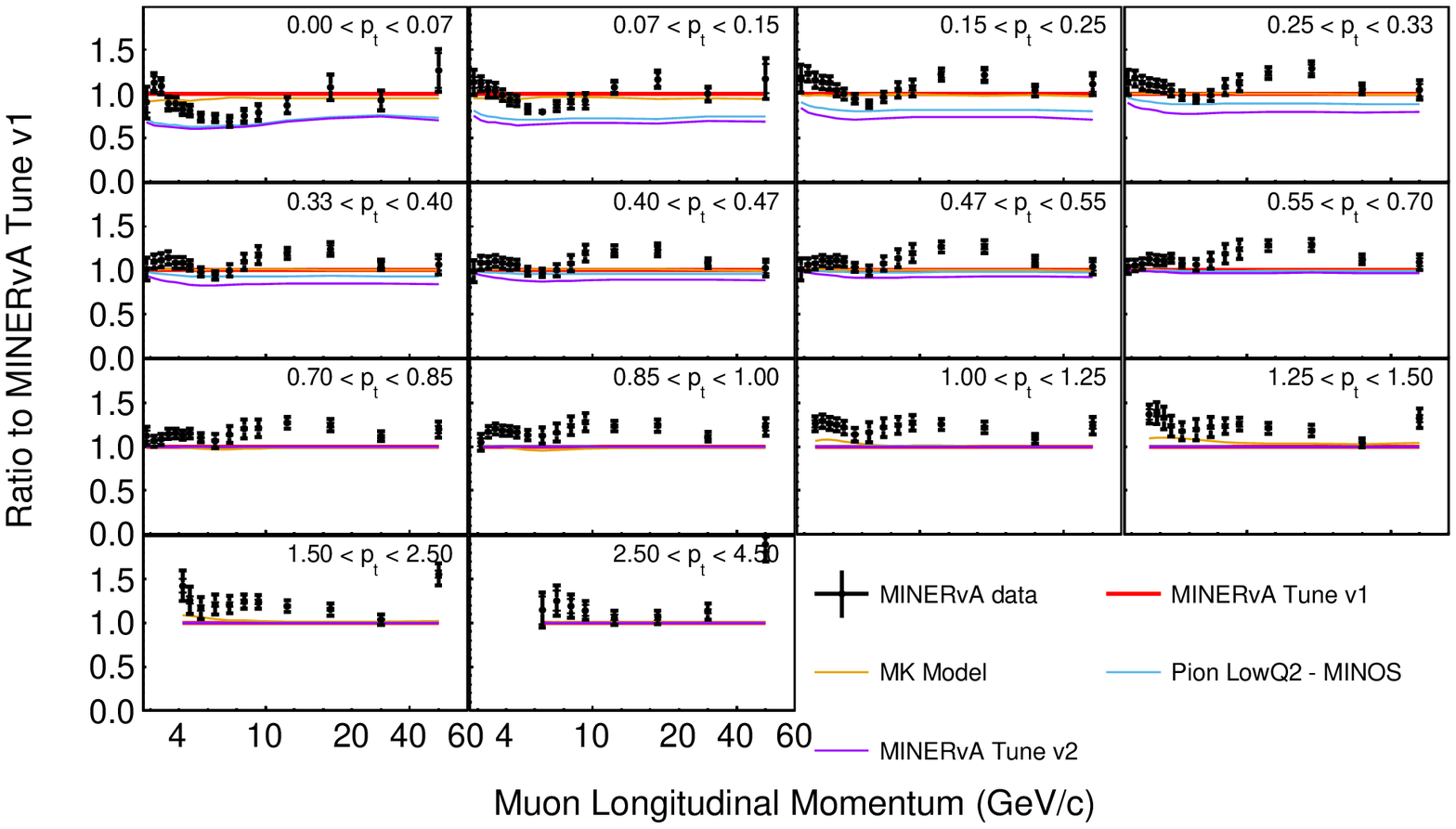}
    \includegraphics[width=\textwidth] {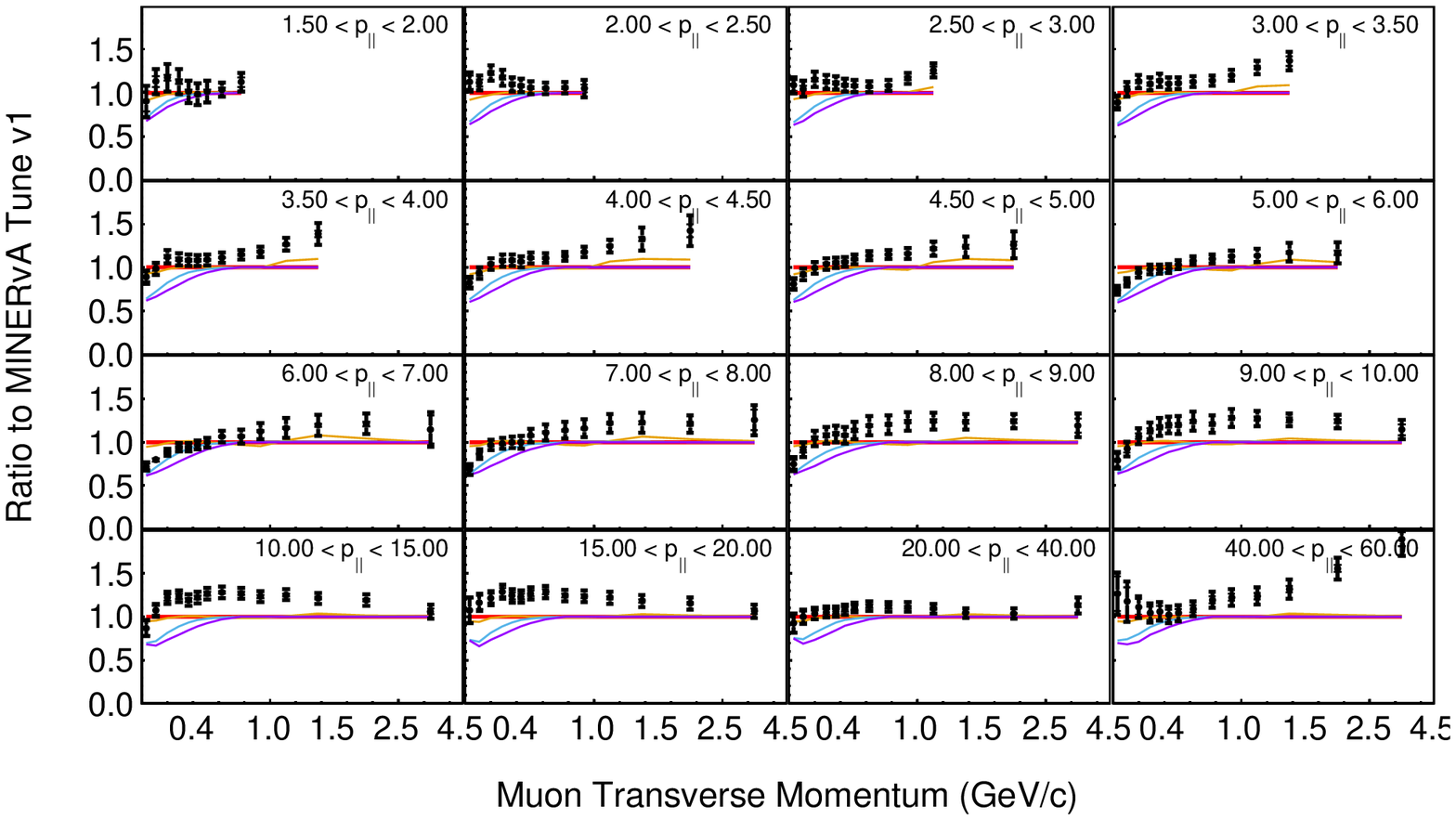}
    \caption{The extracted cross section and predictions modifying the resonant pion prediction taken as a ratio to \tune. Inner (outer) ticks denote statistical (total) uncertainty.}
    \label{fig:set2comp}
\end{figure*}

 The third set of comparisons, see Fig. \ref{fig:set2comp}, use \tune~as the baseline with modifications of various resonant pion production channels. These include:
 \begin{enumerate}[label=(\alph*),noitemsep]
     \item \tune~with the nominal GENIE resonant pion model, (\tune);
     \item Case a) but with the non-resonant pion reduction removed and a reweighting of resonant pion production to the MK model \cite{Kabirnezhad:2017jmf}, (MK Model);
     \item Case a) but with a reweighting to reduce the resonant pion cross section at low $Q^2$ according to a MINOS parameterization \cite{Adamson:2014pgc}, (Pion LowQ2-MINOS);
     \item Case a) but with a reweighting to reduce the resonant pion cross section at low $Q^2$ according to a \minerva~ parameterization \cite{Stowell:2019zsh}, (MINERvA Tune v2).
 \end{enumerate}
 
\begin{figure*}
    \centering
    \includegraphics[width=\textwidth] {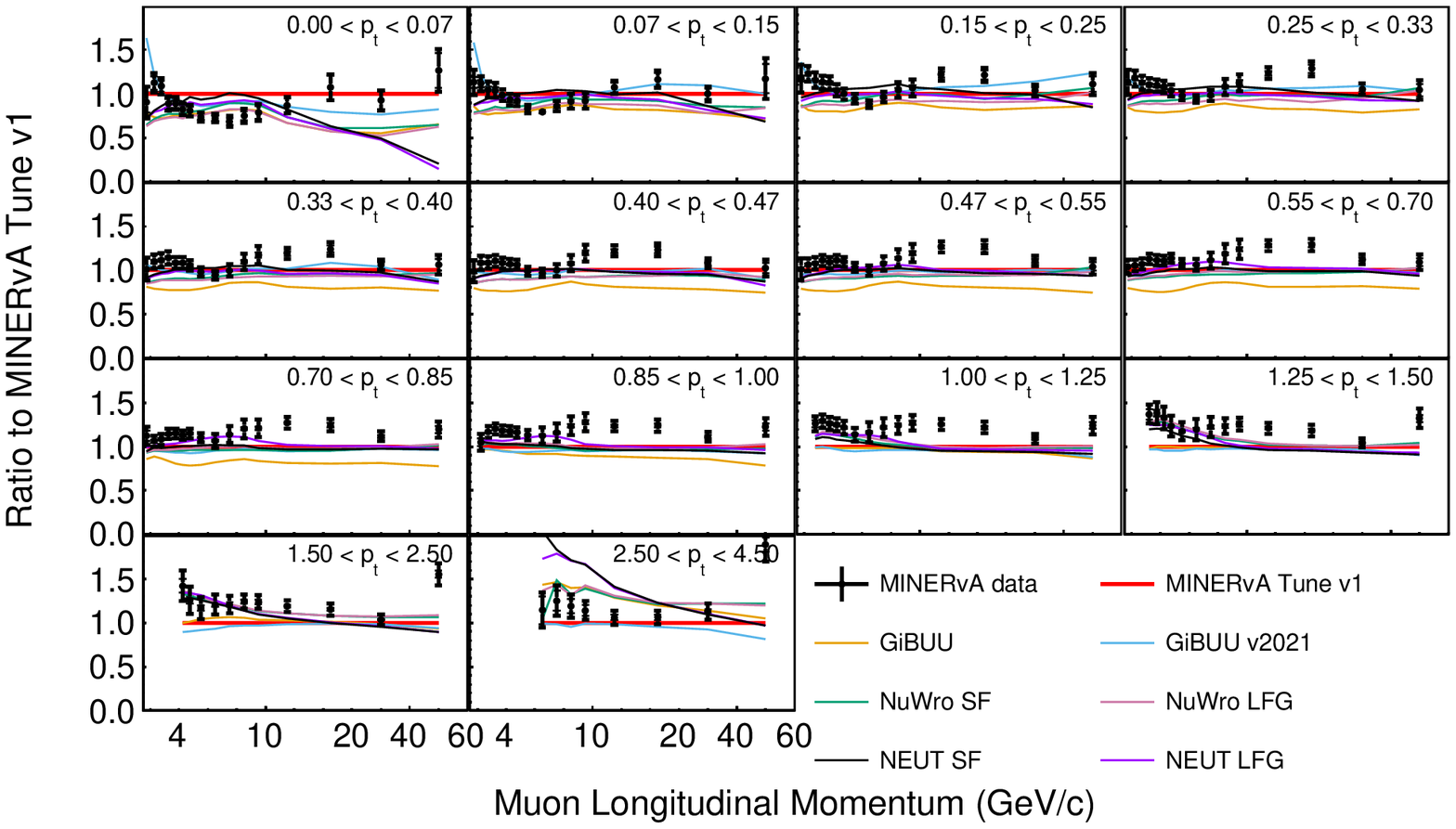}
    \includegraphics[width=\textwidth] {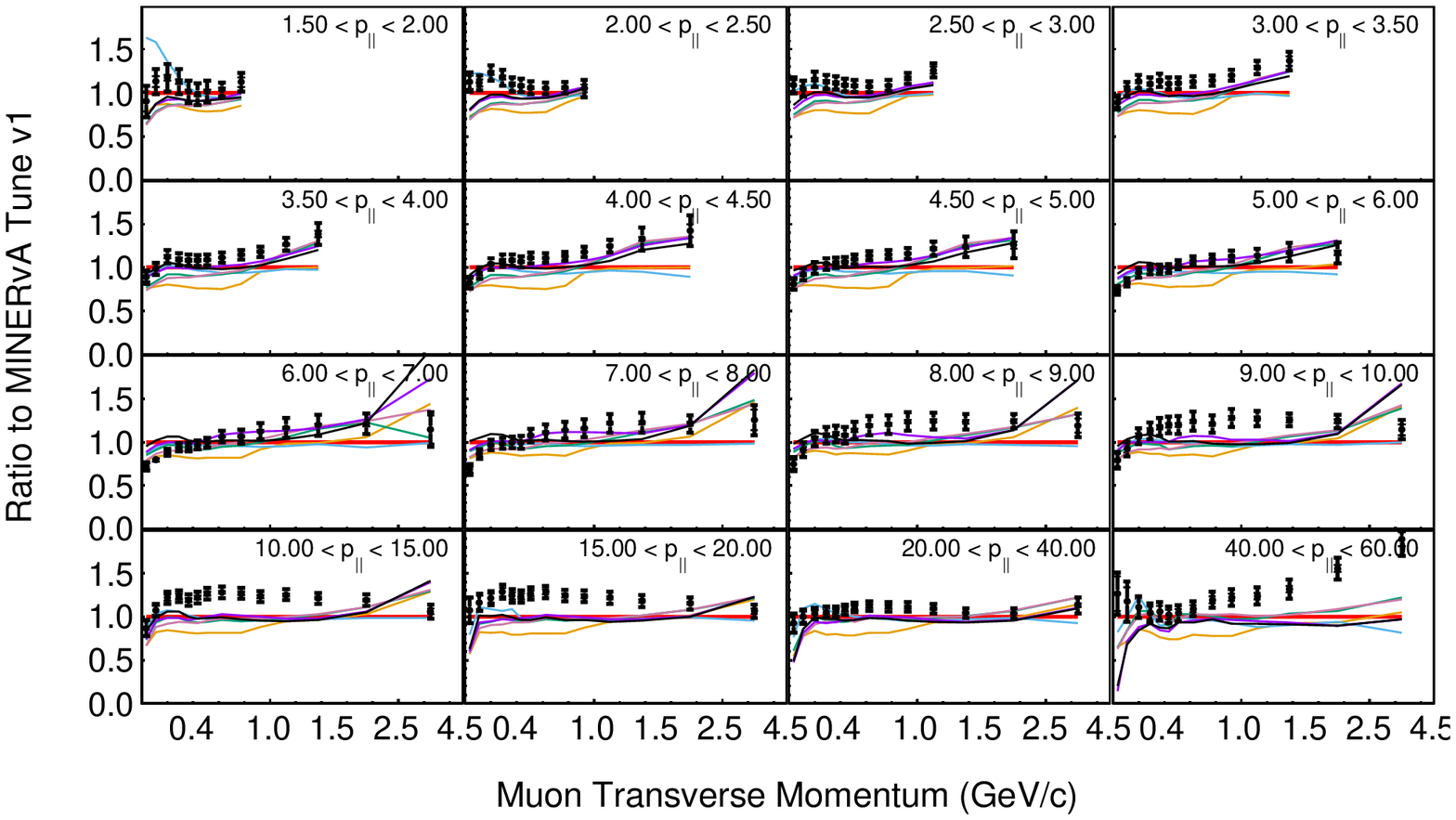}
    \caption{The extracted cross section and external generator predictions displayed as a ratio to \tune. Inner (outer) ticks denote statistical (total) uncertainty.}
    \label{fig:set3comp}
\end{figure*}

Figure~\ref{fig:set3comp} shows a comparison between various neutrino generator predictions including GiBUU~\cite{Buss:2011mx,Gallmeister:2016dnq} for two different versions, two different nuclear models from NuWro~\cite{Golan:2012rfa,Golan:2012wx}, and NEUT~\cite{Hayato:2009zz}.\footnote{Predictions for NuWro and NEUT were produced using NUISANCE \cite{Stowell:2016jfr}}

Table \ref{tab:DDModelComp} gives the $\chi^2$ statistics comparing data to the various model predictions listed above. The $\chi^2$ is calculated using Eq. \ref{eq:chimodel} and summing over $i,j$ which takes into account the covariance between bins. The standard $\chi^2$ calculation assumes the underlying uncertainty is normally distributed. This assumption is not correct as some normally distributed uncertainties manifest as log-normal when propagated to the final result. An example is the flux normalization which introduces uncertainty in the measurement by division. To understand the span of these differences we report both the standard $\chi^2$ and a log-normal version. 
The model with the lowest standard and log-normal $\chi^2$ is the NuWro prediction with a local Fermi gas (LFG) nuclear model, which like the NEUT local Fermi gas prediction, does well at predicting the data between $4<$\pz$<8~GeV$ and \pt$>1.5~GeV$. The effect of Peele's Pertinent Puzzle \cite{BoxCox:1964,Peelle:1987,CARLSON20093215} is clearly shown by the effect of the GiBUU v2019 prediction (or the difference in the peak region between \tune~and GENIE v2.12.6 with Valencia 2p2h in Figs. \ref{fig:1Dptxsec} and \ref{fig:1Dpzxsec}) which, by eye, has a different normalization than the data. This effect occurs when the dominant uncertainty of a result is a highly correlated normalization uncertainty. The log-normal $\chi^2$ increases significantly for models where the dominant difference with data is the normalization. Because of the flux is a dominant uncertainty in this analysis the log-normal $\chi^2$ is a better estimation, but the ordering of models within each group of model modifications is the same using either estimator. 

The \tune~agrees better with the data than GENIE v2.12.6 with Valencia 2p2h. The inclusion or removal of components can either improve or degrade the data Monte Carlo agreement. The inclusion of the non-resonant pion reduction results in poorer agreement than the base model. GENIE v2.12.6 with Valencia 2p2h and QE RPA and a non-resonant pion production reduction is the best combination to predict the data while the model with the low recoil enhancement has the largest $\chi^2$. The \tune~is supported by a variety of exclusive measurements in a 3 GeV neutrino focused beam \cite{Rodrigues:2016xjj,Lu:2018stk,Ruterbories:2018gub} indicate a need for the all modifications but when compared against data using the 6 GeV beam \cite{Carneiro:2019jds} GENIE v2.12.6 with Valencia 2p2h and QE RPA and a non-resonant pion production reduction is less discrepant. The modifications to DIS have mild modifications to the overall prediction. Models which further modify the resonant pion cross section improve the prediction. The data prefer a low $Q^2$ type suppression for resonant pions.  To understand the effect of the modifications in detail the $\chi^2$ is broken down into contributions for each kinematic bin.

\begin{figure*}
    \centering
    \includegraphics[width=0.49\linewidth]{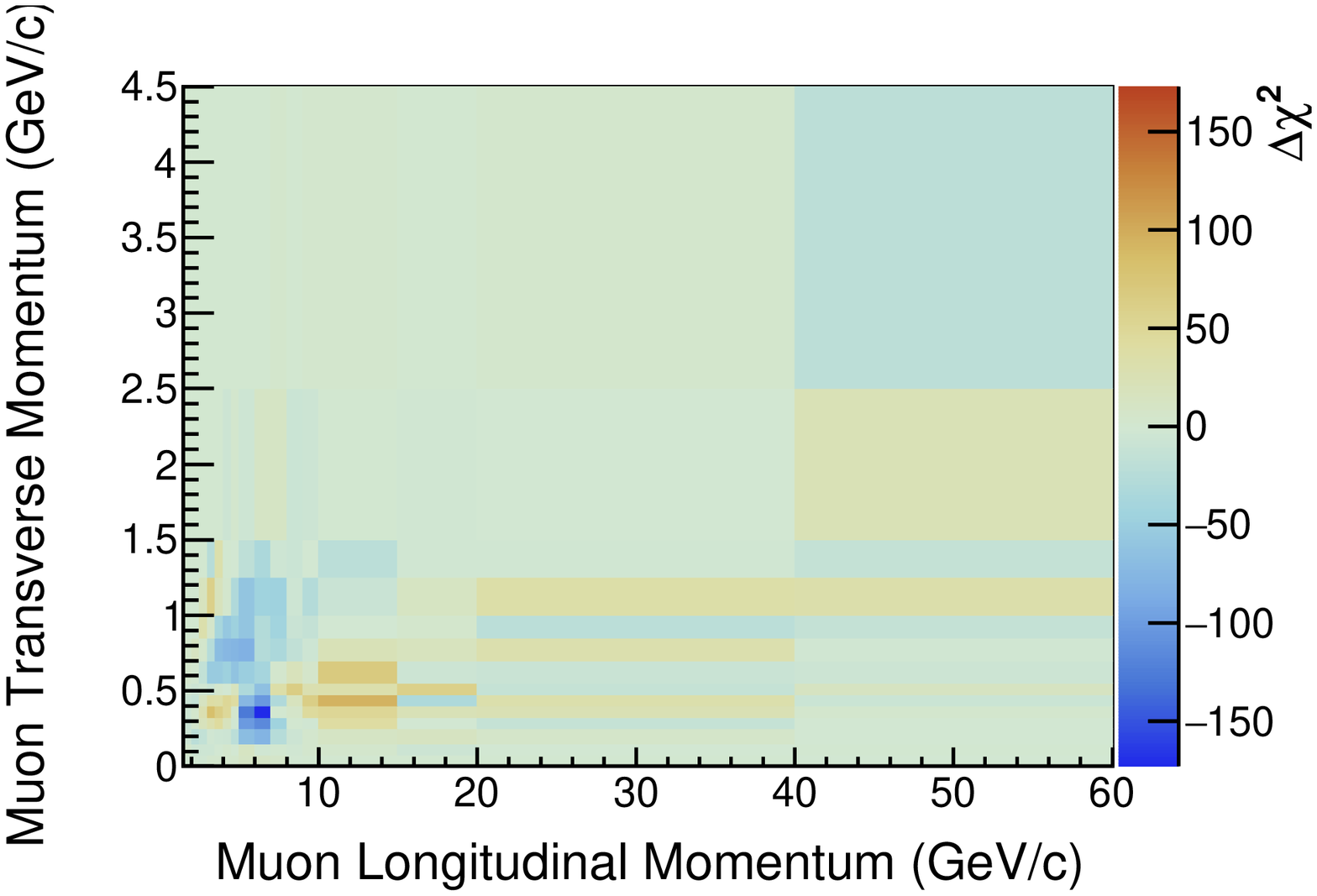}
    \includegraphics[width=0.49\linewidth]{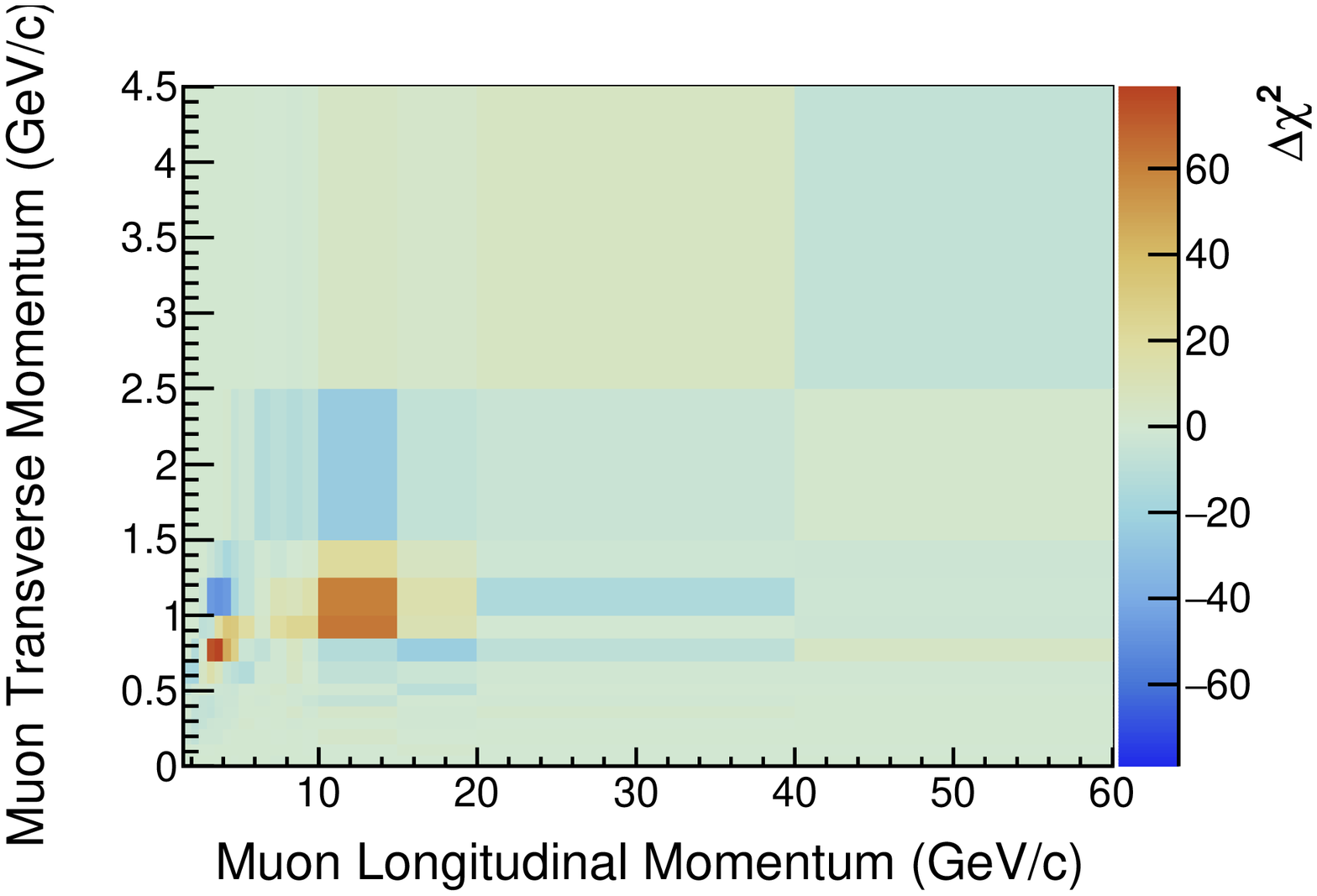}
    \includegraphics[width=0.49\linewidth]{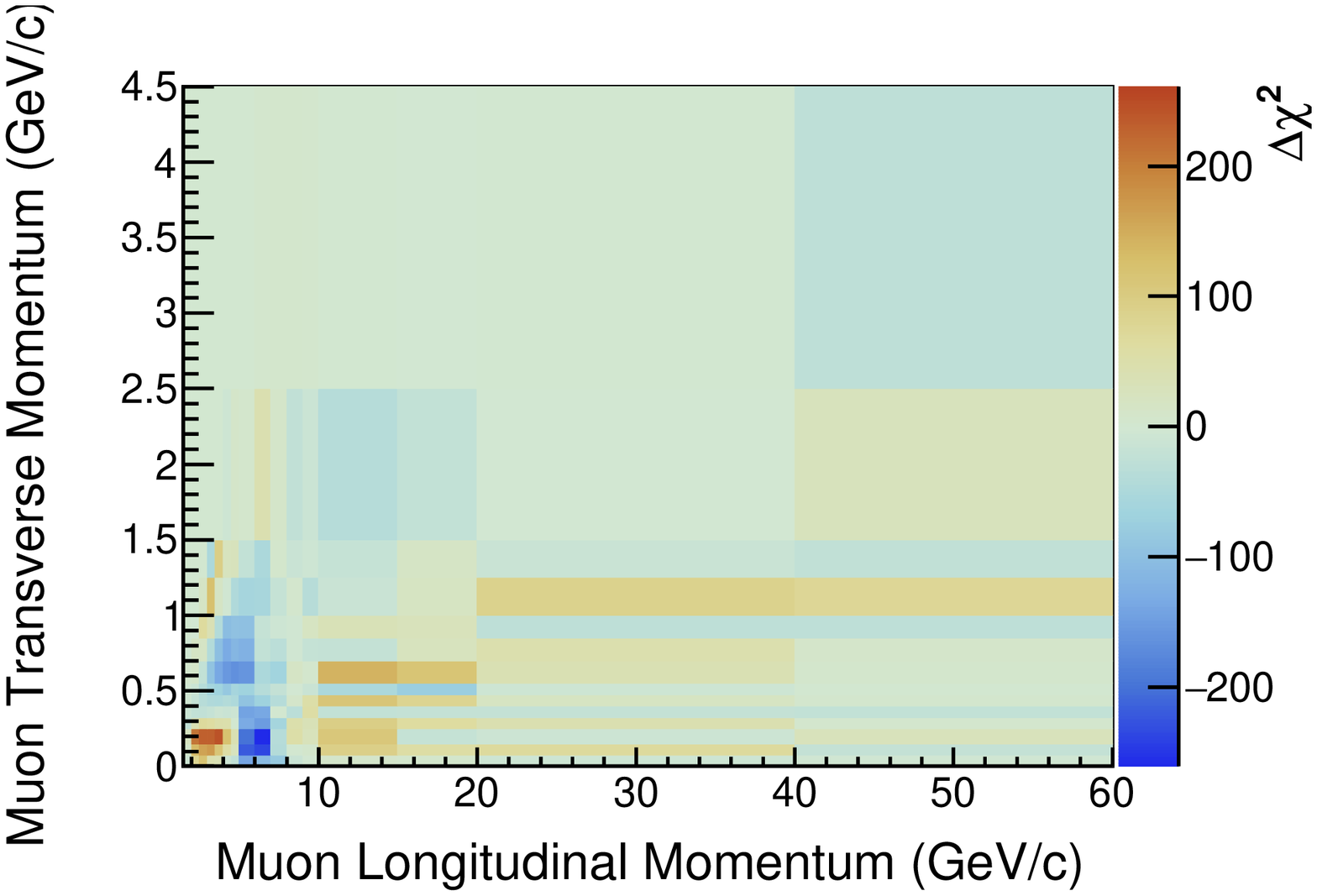}  
    \includegraphics[width=0.49\linewidth]{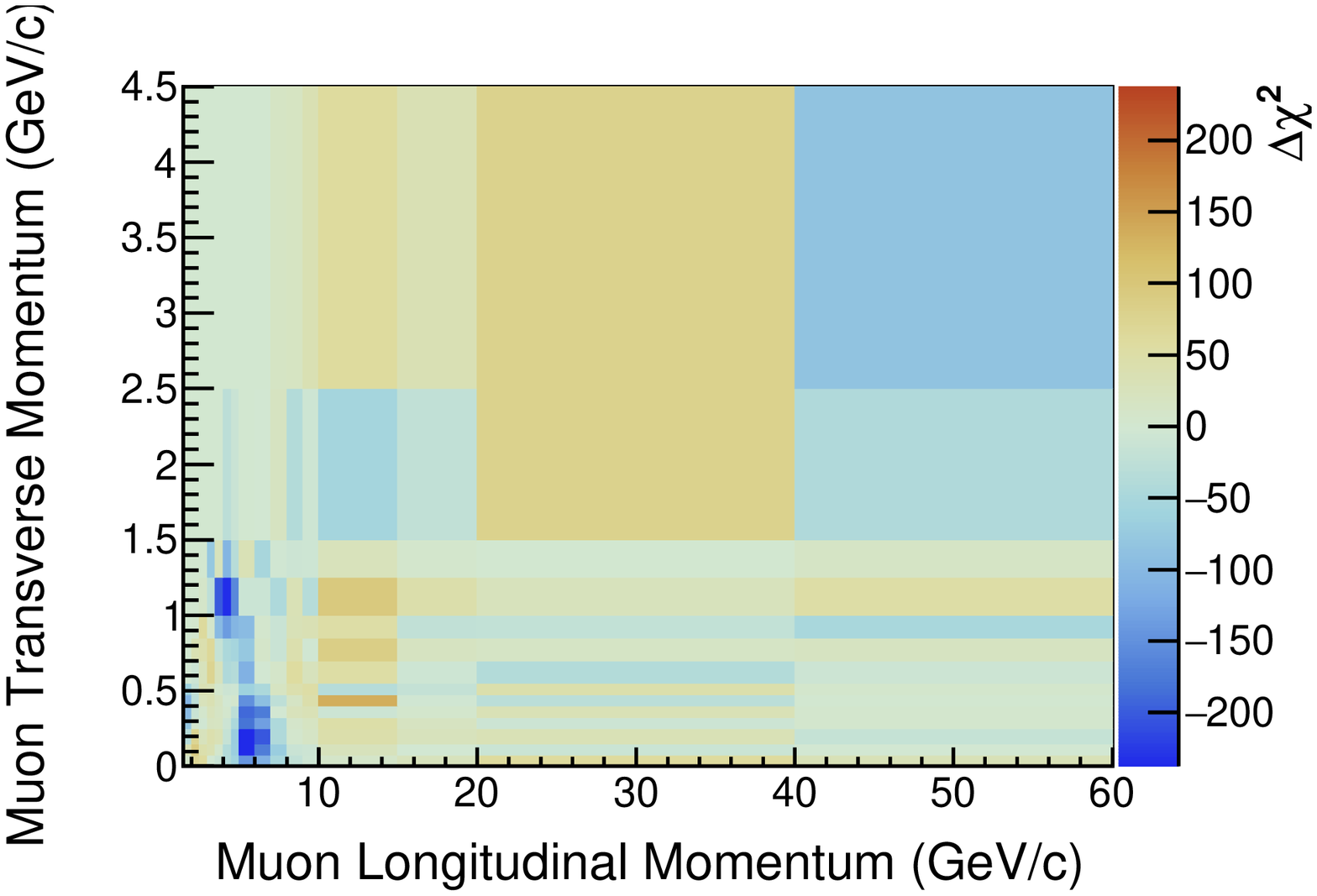}
    \caption{Difference of $\chi^2$ between \tune~and the leading model from each group. Top left is GENIE 2.12.6 and QE RPA and a non-resonant pion reduction; top right is \tune~with nCTEQ$\nu$; bottom left is \tune~with low $Q^2$ pion - MINOS; bottom right is NuWro with local Fermi gas.}
    \label{fig:dchi2models}
\end{figure*}

Figure \ref{fig:dchi2models} shows the bin-by-bin contributions to the overall $\chi^2$. The best prediction, based on overall $\chi^2$, from each model group from Sec. \ref{sec:comparisons} is compared against the \tune. The metric used is the difference in $\chi^2$ on a bin-by-bin basis. The value for the $i^{th}$ bin is the result of the calculation shown in Eqs.~\ref{eq:chimodel} - \ref{eq:dchi2}

\begin{multline}
    \chi^2_{{i,j}_{model}} = (x_{i,measured} - x_{i,expected_{\mathrm{model}}})\times\\ V_{ij}^{-1} \times(x_{j,measured} - x_{j,expected_{\mathrm{model}}}) ,
    \label{eq:chimodel}
\end{multline}
\begin{equation}
    \Delta\chi^2_i = \sum_{j}(\chi^2_{i,j_{\mathrm{model}}}-\chi^2_{i,j_{\mathrm{\tune}}}) ,
    \label{eq:dchi2}
\end{equation}
 where $x$ is the cross section and $V$ is the measurement covariance matrix.
Due to the anti-correlations introduced by the unfolding procedure some large bin-to-bin anti-correlations will appear in this metric. Negative values indicate an improvement due to the model with respect to \tune. 

GENIE 2.12.6 with Valencia 2p2h and QE RPA and a non-resonant pion reduction, \tune~with low $Q^2$ pion suppression models, and NuWro with a local Fermi gas roughly improve in the same regions of phase space. The low $Q^2$ pion suppression modification appears to be an overcorrection for the lowest \pt-\pz~bins, which is evident in Fig. \ref{fig:set2comp}. The modification to the DIS model is different than the other three modifications. The nCTEQ15 model improves agreement in regions of increasing \pt~as a function of \pz. This is a kinematic boundary between the ``Soft DIS" and ``True DIS". 

\begingroup
\squeezetable
\begin{table}[]
\begin{tabular}{lcr}
\hline
Process Variant & Standard $\chi^{2}$ & Log-normal $\chi^{2}$ \\ 
\hline \hline
\tune & 6786 & 7494 \\ \hline
GENIE 2.12.6 & 8241 & 7892 \\ \hline
GENIE 2.12.6 and NonResPionTune Only & 9764 & 9910 \\ \hline 
GENIE 2.12.6 and QE RPA & 5661 & 6544 \\ \hline 
GENIE 2.12.6 and Low Recoil Enhancement & 12345 & 12074 \\ \hline 
\tune~with nCTEQ15 & 6803 & 7530\\\hline 
\tune~with nCTEQ$\nu$ & 6954 & 7762\\\hline 
\tune~with AMU & 7652& 8793\\ \hline 
\tune~using MK& 6224 & 7049\\\hline 
\tune~with \\Low Q$^2$ Pion - MINOS & 4553 & 6388\\\hline
MINERvA GENIE tune v2 &5022 & 7833\\\hline 
GiBUU v2019 & 5800& 9246\\\hline 
GiBUU v2021 & 5594& 6779\\\hline 
NuWro with Spectral Function & 5151 & 6394 \\\hline 
NuWro with Local Fermi Gas & 3789 & 4944\\\hline 
NEUT with Spectral Function & 9151 & 10020 \\\hline 
NEUT with Local Fermi Gas & 6251 & 7452\\\hline 

\hline 
\end{tabular}
\caption{$\chi^{2}$ of various model variants compared to data using the standard and log-normal $\chi^{2}$ where there are 205 degrees of freedom.}
\label{tab:DDModelComp}
\end{table}
\endgroup

\section{Conclusions}
\label{sec:Conclusion}
This paper presents the double-differential and single-differential cross sections as a function of the muon transverse and longitudinal momenta using data recorded by the \minerva~detector in the NuMI beamline. The high statistics and high neutrino energy of these new data, combined with a low flux uncertainty compared to previous measurements~\cite{Filkins:2020xol} mean that new regions of kinematic phase space can be examined to unprecedented precision.  
The data are compared to model predictions that reweight different components in the GENIE prediction or to external generator predictions. Some modifications to the GENIE prediction are inspired by measurements of previous exclusive or restricted phase space measurements on earlier data taken by MINERvA. Other modifications to the GENIE prediction (MK, AMU, nCTEQ15, and nCTEQ$\nu$) represent replacements of a particular set of interaction channels. Finally, generators other than GENIE provide a different set of nuclear models, particle transport, and interaction channel models. None of these predictions describe the data well based on $\chi^2$ tests. Similar models were least discrepant when compared with the inclusive double-differential cross section measured~\cite{Filkins:2020xol} in the 3 GeV neutrino focused beam.

The single- and double- differential measurements provide indication of the need for a low Q$^2$ suppression (low \pt) for the resonant pion production channel. In addition, most of the components in \tune~are favored, while the low recoil enhancement from increased 2p2h production is disfavored. Comparing the bin-by-bin $\chi^2$, in Fig. 19, the application of the least discrepant DIS model (nCTEQ15) indicates large changes in the region between ``Soft DIS'' and ``True DIS''. Overall, NuWro with a local Fermi gas nuclear model best describes the data. The result cannot differentiate the specific source of mismodeling in regions where all the underlying processes contribute to the prediction. Other methods, either via exclusive, semi-inclusive, or inclusive measurements using other kinematic variables are needed to investigate these complex regions.  This work represents an important benchmark that can be used to 
validate future ensembles of models tuned to agree with exclusive results.  

\begin{acknowledgments}

This document was prepared by members of the MINERvA Collaboration using the resources of the Fermi National Accelerator Laboratory (Fermilab), a U.S. Department of Energy, Office of Science, HEP User Facility. Fermilab is managed by Fermi Research Alliance, LLC (FRA), acting under Contract No. DE-AC02-07CH11359.
These resources included support for the MINERvA construction project, and support
for construction also
was granted by the United States National Science Foundation under
Award No. PHY-0619727 and by the University of Rochester. Support for
participating scientists was provided by NSF and DOE (USA); by CAPES
and CNPq (Brazil); by CoNaCyT (Mexico); by Proyecto Basal FB 0821, CONICYT PIA ACT1413, and Fondecyt 3170845 and 11130133 (Chile); 
by CONCYTEC (Consejo Nacional de Ciencia, Tecnolog\'ia e Innovaci\'on Tecnol\'ogica), DGI-PUCP (Direcci\'on de Gesti\'on de la Investigaci\'on  - Pontificia Universidad Cat\'olica del Peru), and VRI-UNI (Vice-Rectorate for Research of National University of Engineering) (Peru); NCN Opus Grant No. 2016/21/B/ST2/01092 (Poland); by Science and Technology Facilities Council (UK)); by EU Horizon 2020 Marie Skłodowska-Curie Action; by a Cottrell Postdoctoral Fellowship, Research Corporation for Scientific Advancement award number 27467 and National Science Foundation Award CHE2039044  We thank the MINOS Collaboration for use of its near detector data. Finally, we thank the staff of
Fermilab for support of the beam line, the detector, and computing infrastructure.

%
%
%
%
%
%
%
%
%
%
%
%
\end{acknowledgments}
\clearpage

\bibliography{main}

\end{document}